\documentclass[10pt]{article}
 \usepackage{amssymb, amsmath}

\usepackage{amsmath}
\usepackage{latexsym}
\usepackage{amssymb}

\usepackage{amsthm}

\date{}

\font\tengoth=eufm10 at 10pt
\font\sevengoth=eufm7 at 6pt
\newfam\gothfam
\textfont\gothfam=\tengoth
\scriptfont\gothfam=\sevengoth

\newcommand{\mlabel}[1]{\marginpar{#1}\label{#1}}

\renewcommand{\:}{\colon}
\newcommand{\1}{\mathbf{1}}

\newcommand{\cA}{\mathcal{A}}

\newcommand{\cD}{\mathcal{D}}
\newcommand{\cE}{\mathcal{E}}
\newcommand{\cF}{\mathcal{F}}

\newcommand{\cH}{\mathcal{H}}
\newcommand{\cK}{\mathcal{K}}
\newcommand{\cL}{\mathcal{L}}

\newcommand{\cN}{\mathcal{N}}
\newcommand{\cO}{\mathcal{O}}

\newcommand{\cS}{\mathcal{S}}

\newcommand{\subeq}{\subseteq}
\newcommand{\supeq}{\supseteq}

\newcommand{\into}{\hookrightarrow}
\newcommand{\eps}{\varepsilon}

\newcommand{\shalf}{{\textstyle{\frac{1}{2}}}}

\newcommand{\Z}{{\mathbb Z}}
\newcommand{\R}{{\mathbb R}}
\newcommand{\C}{{\mathbb C}}

\newcommand{\T}{{\mathbb T}}
\newcommand{\V}{{\mathbb V}}

\newcommand{\bH}{{\mathbb H}}
\newcommand{\bS}{{\mathbb S}}

\newcommand{\bV}{{\mathbb V}}

\renewcommand{\hat}{\widehat}

\renewcommand{\tilde}{\widetilde}

\newcommand{\SO}{\mathop{{\rm SO}}\nolimits}

\newcommand{\OO}{\mathop{\rm O{}}\nolimits}

\newcommand{\U}{\mathop{\rm U{}}\nolimits}

\newcommand{\Fix}{\mathop{{\rm Fix}}\nolimits}

\renewcommand{\Re}{\mathop{{\rm Re}}\nolimits}
\renewcommand{\Im}{\mathop{{\rm Im}}\nolimits}

\newcommand{\Hom}{\mathop{{\rm Hom}}\nolimits}

\newcommand{\Herm}{\mathop{{\rm Herm}}\nolimits}

\newcommand{\Aut}{\mathop{{\rm Aut}}\nolimits}

\newcommand{\id}{\mathop{{\rm id}}\nolimits}

\newcommand{\im}{\mathop{{\rm im}}\nolimits}

\newcommand{\supp}{\mathop{{\rm supp}}\nolimits}

\newcommand{\Spann}{\mathop{{\rm span}}\nolimits}

\newcommand{\Bil}{\mathop{{\rm Bil}}\nolimits}
\newcommand{\Sesq}{\mathop{{\rm Sesq}}\nolimits}

\renewcommand{\phi}{\varphi}

\newcommand{\Rarrow}{\Rightarrow}
 
\newcommand{\oline}{\overline}

\newcommand{\la}{\langle}
\newcommand{\ra}{\rangle}

\newcommand{\res}{\vert}

\newcommand{\Spec}{{\rm Spec}}

\newcommand{\ssssarr}{\hbox to 15pt{\rightarrowfill}}
\newcommand{\sssarr}{\hbox to 20pt{\rightarrowfill}}
\newcommand{\ssarr}{\hbox to 30pt{\rightarrowfill}}
\newcommand{\sarr}{\hbox to 40pt{\rightarrowfill}}
\newcommand{\arr}{\hbox to 60pt{\rightarrowfill}}
\newcommand{\larr}{\hbox to 60pt{\leftarrowfill}}
\newcommand{\Arr}{\hbox to 80pt{\rightarrowfill}}

\def\theoremname{Theorem}
\def\propositionname{Proposition}
\def\corollaryname{Corollary}
\def\lemmaname{Lemma}
\def\remarkname{Remark}
\def\conjecturename{Conjecture} 

\def\definitionname{Definition}
\def\exercisename{Exercise}
\def\examplename{Example}
\def\examplesname{Examples}
\def\problemname{Problem}
\def\problemsname{Problems}

\def\satzname{Satz} 
\def\koroname{Korollar}
\def\folgname{Folgerung}
\def\bemerkname{Bemerkung}
\def\aufgname{Aufgabe}

\def\beisname{Beispiel}
\def\beissname{Beispiele}
\def\bewname{Beweis}

\def\@thmcounter#1{\noexpand\arabic{#1}}
\def\@thmcountersep{}
\def\@begintheorem#1#2{\it \trivlist \item[\hskip 
\labelsep{\bf #1\ #2.\quad}]}
\def\@opargbegintheorem#1#2#3{\it \trivlist
      \item[\hskip \labelsep{\bf #1\ #2.\quad{\rm #3}}]}
\makeatother
\newtheorem{theor}{\theoremname}[section]
\newtheorem{propo}[theor]{\propositionname}
\newtheorem{coro}[theor]{\corollaryname}
\newtheorem{lemm}[theor]{\lemmaname}

\newenvironment{thm}{\begin{theor}\it}{\end{theor}}

\newenvironment{prop}{\begin{propo}\it}{\end{propo}}

\newenvironment{cor}{\begin{coro}\it}{\end{coro}}

\newenvironment{lem}{\begin{lemm}\it}{\end{lemm}}

\newtheorem{rema}[theor]{\remarkname}

\newenvironment{rem}{\begin{rema}\rm}{\end{rema}}

\newtheorem{stepnow}[theor]{}

\newtheorem{defin}[theor]{\definitionname} 

\newenvironment{defn}{\begin{defin}\rm}{\end{defin}}

\newtheorem{exerc}{\exercisename}[section]

\newtheorem{exa}[theor]{\examplename}

\newenvironment{ex}{\begin{exa}\rm}{\end{exa}}

\newtheorem{exas}[theor]{\examplesname}

\newtheorem{conj}[theor]{\conjecturename}

\newtheorem{pro}[theor]{\problemname}

\newenvironment{prob}{\begin{pro}\rm}{\end{pro}}

\newtheorem{prs}[theor]{\problemsname}

\newtheorem{aufg}{\aufgname}[section]

\newenvironment{prf}{\begin{proof}}{\end{proof}}
 
\newcommand{\pmat}[1]{\begin{pmatrix} #1 \end{pmatrix}}

{\hfill\qed\end{trivlist}}

\newenvironment{beweis*}{\begin{trivlist}\item[\hskip%
\labelsep{\bf\bewname.\quad}]}%
{\end{trivlist}}

\newtheorem{satzn}[theor]{\satzname}

\newtheorem{koro}[theor]{\koroname}

\newtheorem{folg}[theor]{\folgname}

\newtheorem{bem}[theor]{\bemerkname}

\newtheorem{aufgn}[theor]{\aufgname}

\newtheorem{beis}[theor]{\beisname}

\newtheorem{beiss}[theor]{\beissname}

\renewcommand{\phi}{\varphi}
\renewcommand{\mlabel}{\label} 
\newcommand\bv{{\bf{v}}}
\newcommand\bw{{\bf{w}}}

\begin{document} 


\title{KMS conditions, standard real subspaces \\  and reflection positivity on 
the circle group} 
\author{Karl-Hermann Neeb, Gestur \'Olafsson}

\maketitle

\begin{abstract}
In the present paper we continue our investigations of the 
representation theoretic side of reflection positivity by studying 
positive definite functions $\psi$ on the additive group $(\R,+)$ 
satisfying a suitably defined KMS condition. 
These functions take values in the space $\Bil(V)$ of 
bilinear forms on a real vector space~$V$. As in quantum statistical 
mechanics, the KMS condition is defined in terms of an analytic continuation 
of $\psi$ to the strip $\{ z \in \C \: 0 \leq \Im z \leq \beta\}$ 
with a coupling condition $\psi(i\beta + t) 
= \oline{\psi(t)}$ on the boundary. 
Our first main result consists of a characterization 
of these functions in terms of modular objects $(\Delta, J)$ 
($J$~an antilinear involution and $\Delta > 0$ selfadjoint 
with $J\Delta J = \Delta^{-1}$) and an integral representation. 

Our second main result is the existence 
of a $\Bil(V)$-valued positive definite function $f$ on the group 
$\R_\tau = \R \rtimes \{\id_\R,\tau\}$ with $\tau(t) = -t$ 
satisfying $f(t,\tau) = \psi(it)$ for $t \in \R$. We thus obtain a 
$2\beta$-periodic 
unitary one-parameter group on  the GNS space $\cH_f$ for which the 
one-parameter group on  the GNS space $\cH_\psi$ is obtained by 
Osterwalder--Schrader quantization. 

Finally, we show that the building blocks of these representations 
arise from bundle-valued Sobolev spaces corresponding to the kernels 
$(\lambda^2 - \frac{d^2}{dt^2})^{-1}$ on the circle $\R/\beta\Z$ 
of length $\beta$.
\end{abstract}

\section{Introduction} 

In this note we continue our investigations of the mathematical 
foundations of \textit{reflection positivity}, a basic concept in constructive quantum 
field theory (\cite{GJ81, KL83, JOl98, JOl00, AFG86, JR07}).  
Originally, reflection positivity, also called Osterwalder--Schrader positivity, 
arises as a requirement on the euclidean side to establish a 
duality between euclidean and relativistic quantum field theories (\cite{OS73}). 
It is closely related to ``Wick rotation'' or 
``analytic  continuation'' in the time variable 
from the real  to the imaginary axis. 

The underlying fundamental concept is that of a 
{\it reflection positive Hilbert space}, introduced in \cite{NO14a}. 
This is a triple $(\cE,\cE_+,\theta)$, 
where $\cE$ is a Hilbert space, $\theta : \cE \to \cE$ is a unitary involution
and $\cE_+$ is a closed subspace of $\cE$ which is $\theta$-positive in the sense that 
$\la \theta v, v \ra \geq 0$ for~$v\in\cE_+$. 

In \cite{NO14a} we introduced the concept of a reflection positive cyclic 
representation $(\pi, \cE, v)$, where $(\cE,\cE_+,\theta)$ is a reflection 
positive Hilbert space and $v \in \cE$ a $\theta$-fixed vector 
(or, more generally, a distribution vector). 
In the present paper we shall see that, 
to treat reflection positive representations of the circle 
group $G = \T$ corresponding to unitary representations 
of the dual group $G^c \cong \R$ arising from 
KMS states, resp., from their modular objects $(\Delta, J)$,\begin{footnote}
{Recall that KMS stands for Kubo--Martin--Schwinger; see cite[\S 5.3.1]{BR96} 
for more on KMS states and their interpretation in Quantum Statistical 
Mechanics as thermal equilibrium states.}
\end{footnote}
we are forced to work in a more general framework, where
the representations are generated by the image 
of an $\R$-linear map $j \: V \to \cE$ from a real vector space $V$ into 
the representation space $\cE$ and where $j(V)$ does not consist of $\theta$-fixed 
vectors. 

To explain the corresponding concept of a {\it reflection positive representation}, 
we start with a {\it symmetric Lie group}, i.e., a pair $(G,\tau)$, 
where $\tau \in \Aut(G)$ is an involution. Then we form the extended group 
$G_\tau :=G \rtimes \{\1,\tau\}$. 
Let $(U,\cE)$ be a unitary representation of $G_\tau$ and let 
$j \: V \to \cE$ be a linear map from the real vector space $V$ to $\cE$. 
Then $(U,\cE,j,V)$ is called {\it reflection positive with respect to 
a subset $G_+\subeq G$} 
if the closed subspace $\cE_+$ generated by $U_{G_+}^{-1}j(V)$ defines a 
reflection positive Hilbert space $(\cE,\cE_+,U_\tau)$. 
Generalizing the well-known Gelfand--Naimark--Segal (GNS) construction  
leads to an encoding of representations generated by $j(V)$ in terms 
of form-valued positive definite functions 
$\psi(g)(v,w) := \la j(v), U_g j(w)\ra$ (\cite{NO15}). 

This paper continues the investigations started in \cite{NO15}, 
where we studied reflection positive representations of the circle 
group and their connections to KMS states, which was largely 
motivated by the work of Klein and Landau in \cite{KL81} 
(see also \cite{CMV01}). 
A long term goal of this project is to combine our representation theoretic 
approach to reflection positivity 
with KMS states of operator algebras and Borchers 
triples corresponding to modular inclusions 
(\cite{BLS11}, \cite{Bo92},\cite{Lo08}, \cite{Sch99}). 

A crucial step in this direction is the 
concept of a positive definite function satisfying a KMS condition 
that can be formulated as follows: First, let $V$ be a real vector space 
and $\Bil(V)$ be the space of real bilinear maps $V \times V \to \C$.
A function $\psi \: \R \to \Bil(V)$ is said to be {\it positive 
definite} if the kernel 
$\psi(t-s)(v,w)$ on $\R \times V$ is positive definite. 
For $\beta > 0$, we consider the open strip 
$\cS_\beta := \{ z \in \C \: 0 < \Im z < \beta\}$. 
We say that a positive definite 
function $\psi \: \R \to \Bil(V)$ satisfies the {\it KMS condition} 
for $\beta > 0$ if $\psi$ 
extends to a function $\oline{\cS_\beta} \to \Bil(V)$ which is pointwise 
continuous and pointwise holomorphic on the interior $\cS_\beta$, and satisfies 
\[ \psi(i \beta+t) = \oline{\psi(t)}\quad \mbox{ for } \quad t \in \R.\] 

The central idea in the classification of positive definite functions 
satisfying a KMS condition is 
to relate them to {\it standard real subspaces} of a (complex) Hilbert space; 
these are closed real subspaces $V \subeq \cH$ for which 
$V \cap i V = \{0\}$ and $V + i V$ is dense 
(cf.\ Definition~\ref{def:1.5}). Any such subspace determines a pair 
$(\Delta, J)$ of {\it modular objects}, where $\Delta$ is a positive selfadjoint 
operator and $J$ an antilinear involution satisfying $J\Delta J = \Delta^{-1}$. 
The connection is established by 
$V = \Fix(J\Delta^{1/2}) = \{ v \in \cD(\Delta^{1/2}) \: J\Delta^{1/2} v = v \}$ 
Our first main result is the following characterization of the 
KMS condition in terms of standard real subspaces. 
Here we write $\Bil^+(V) \subeq \Bil(V)$ for the convex cone of all those 
bilinear forms $f$ for which the sesquilinear extension to $V_\C \times V_\C$ 
is positive semidefinite. 

\begin{thm} {\rm(Characterization of the KMS condition)} \mlabel{thm:1-intro}
Let $V$ be a real vector space and 
let $\psi \:  \R \to \Bil(V)$ be a pointwise continuous positive definite function. 
Then the following are equivalent: 
\begin{itemize}
\item[\rm(i)] $\psi$  satisfies the 
KMS condition for $\beta > 0$. 
\item[\rm(ii)] There exists a standard real subspace 
$V_1$ in a Hilbert space $\cH$ and a linear map 
$j \: V \to V_1$ such that 
\begin{equation}
  \label{eq:pdform}
 \psi(t)(v,w) = \la  j(v), \Delta^{-it/\beta} j(w) \ra \quad \mbox{ for } \quad 
t \in \R,v,w \in V.
\end{equation}
\item[\rm(iii)] There exists a $\Bil^+(V)$-valued regular Borel measure $\mu$ 
on $\R$ satisfying 
\[ \psi(t) = \int_\R e^{it\lambda}\, d\mu(\lambda), 
\quad \mbox{ where  } \quad 
d\mu(-\lambda) = e^{-\beta\lambda}d\oline\mu(\lambda).\] 
\end{itemize}
If these conditions are satisfied, then the function 
$\psi \: \oline{\cS_\beta} \to \Bil(V)$ is pointwise bounded. 
\end{thm}

The equivalence of (i) and (ii) in Theorem~\ref{thm:1-intro}
 describes the tight connection 
between the KMS condition and the modular objects associated to a standard 
real subspace. Part (iii) provides an integral representation that can be 
viewed as a classification result. 

For a function $\psi$ satisfying the $\beta$-KMS condition, analytic 
continuation leads to the operator-valued function  
\[ \phi \: [0,\beta] \to B(V_\C), \qquad 
\la v, \phi(t)w \ra = \psi(it)(v,w).\] 
This function satisfies $\phi(\beta) = \oline{\phi(0)}$, hence 
extends uniquely to a (weak operator) continuous 
function $\phi \: \R \to B(V_\C)$ satisfying 
\begin{equation}
  \label{eq:phi-per}
\phi(t + \beta) = \oline{\phi(t)} \quad \mbox{ for } \quad t \in \R.
\end{equation}

Recall the group $\R_\tau := \R \rtimes \{1,\tau\}$ with 
$\tau(t) = -t$. In Theorem~\ref{thm:2} we 
show that there exists a positive definite function 
\[ f \: \R_\tau \to \Bil(V) 
\quad \mbox{ satisfying } \quad 
f(t,\tau) = \hat\phi(t). \] 
The function $f$ is $2\beta$-periodic,  
hence factors through a function on $\T_{2\beta,\tau} 
:= \R_\tau/\Z 2\beta \cong \OO_2(\R)$. 
This leads to a natural ``euclidean'' counterpart of the 
unitary one-parameter group $U_t = \Delta^{-it/\beta}$ 
associated to the KMS positive definite function~$\psi$. 
To understand the structure of the so arising positive definite 
functions and the corresponding unitary representations 
of $\T_{2\beta,\tau}$, we write 
$f = f_+ + f_-$ with 
$f_+(\beta + t,\tau^\eps) = f_+(t,\tau^\eps)$ 
(the bosonic part) and $f_-(\beta + t,\tau^\eps) = -f_-(t,\tau^\eps)$ 
(the fermionic part). 
Then $f_\pm$ are both positive definite and combine to a matrix valued
positive definite function 
\[ f^\sharp := \pmat{ f_+ &  0 \\ 0 & f_-} \: \R_\tau \to M_2(B(V_\C)) \cong B(V_\C^2)\] 
(Lemma~\ref{lem:indrep}). 
Neglecting an additive summand which is constant, we can now defined 
a unitary representation of the subgroup $P := (\Z \beta)_\tau$ on $V_\C^2$ by 
\[ \rho(\beta,\1) := \pmat{ \1 & 0 \\ 0 & -\1} \quad \mbox{ and } \quad 
\rho(0,\tau) := \pmat{ \1 & 0 \\ 0 & i I},\] 
where $I$ is a complex structure on~$V$.
Then we have the relation 
\[ f^\sharp(hg) = \rho(h) f^\sharp(g) \quad \mbox{ for } \quad 
h \in P, g \in \R_\tau \] 
which determines in particular how $f^\sharp$ is obtained from the function~$\phi$ 
from above. 

For the special case where the real representation corresponding to $\psi$ is isotypic, 
resp., the associated modular operator $\Delta$ is a multiple of the identity, 
the GNS representation $(U^{f^\sharp}, \cH_{f^\sharp})$ 
can be realized on the Hilbert space completion of 
\[ \Gamma_\rho := \{ s \in C^\infty(\R_\tau,V_\C^2) \: (\forall g \in 
\R_\tau, h \in P)\ s(hg) =\rho(h)s(g)\} \] 
with respect to the scalar product 
\[ \la s_1, s_2 \ra := \frac{1}{2\beta}\int_0^{2\beta}
\la s_1(t,\1), ((\lambda^2 - \Delta)^{-1} s_2)(t,\1) \ra \, dt, 
\quad\mbox{ where } \quad 
\Delta = \frac{d^2}{dt^2}. \] 
On this space $\R_\tau$ acts by right translation. 
This provides a natural ``euclidean realization'' of our representation 
on the Riemannian manifold $\T_\beta \cong \bS^1$ in the spirit 
of \cite{AFG86, Di04, JR07}. 

We conclude this paper with a short Section~\ref{sec:5}, in which 
we prove a version of Theorem~\ref{thm:1-intro} for $\beta = \infty$ 
which connects naturally to our previous work on dialations of semigroups 
of contractions 
in~\cite{NO14b}. In two appendices we provide some background material. 
Appendix~\ref{sec:a} recalls some facts on positive definite kernels and 
discusses in particular the connection between complex and real-valued kernels. 
Appendix~\ref{sec:b} discusses standard real subspaces in terms 
of skew-symmetric contractions on real Hilbert spaces. This perspective 
was crucial for the present paper, and we expect it to be 
useful in other contexts as well. 

In a subsequent paper \cite{NO16} we extend the results 
obtained here for the group 
$G = \OO_2(\R) = \SO_2(\R)_\tau$ to more general groups such as 
$\OO_{n+1}(\R)$ (where reflection positivity refers to 
the sphere $\bS^{n}$) and  $\OO_{1,n}(\R)$ (where reflection positivity 
refers to the $n$-dimensional hyperbolic space $\bH^n$). 
Eventually, we would like to see how our representation theoretic 
analysis can be blended with the existing work on 
relativistic KMS conditions \cite{BB94, GJ06} and in particular with~\cite{BJM13, 
BJM15}. \\

The close connection between modular objects $(\Delta, J)$ 
and standard real subspaces was first explored by Rieffel and van Daele 
in \cite{RvD77}. They also define a notion of a KMS condition 
for a unitary one-parameter group $(U_t)_{t \in \R}$ on a complex Hilbert space $\cH$ 
with a real subspace~$V \subeq \cH$. In our terms, their condition means that 
the function $\psi \:\R \to \Bil(V), \psi(t) = \la v, U_t w \ra$ satisfies
the KMS condition for $\beta = -1$ (which refers to a function on the strip 
$\{ -1 < \Im z < 0\}$). 
>From \cite[Prop.~3.7]{RvD77} one can easily derive the implication (ii) $\Rarrow$ (i) 
of Theorem~\ref{thm:1-intro} (cf.\ also \cite[Prop.~3.7]{Lo08}). In this case 
\cite[Thm.~3.8]{RvD77} even implies that $U_t = \Delta^{-it/\beta}$ is 
the unique unitary one-parameter group satisfying the KMS condition for~$\beta$.
>From \cite[Thm.~3.9]{RvD77} one can also derive the implication (i) $\Rarrow$ (ii). 
Instead of $\Delta$, Rieffel and van Daele work with the bounded 
operator $R = 2(\1 + \Delta)^{-1}$ which is the sum of the orthogonal 
projections of the real Hilbert space $\cH$ onto the closed subspaces 
$V$ and $iV$. In our context this operator appears as 
$\1 + i \hat C$ for the skew-hermitian operator 
$\hat C = i \frac{\Delta - \1}{\Delta + \1}$ (Lemma~\ref{lem:3.3}). 

In the context of free fields, the interplay between 
standard real subspaces and von Neumann algebras of operators 
on Fock space has already been studied by 
Araki \cite{Ar63} and Eckmann/Osterwalder \cite{EO73}. 
We refer to \cite{Yn94} for some particularly interesting concrete 
subspaces corresponding to fields on light rays 
and to \cite{Ra00} for descriptions of standard real subspaces 
in terms of boundary values of holomorphic functions.
Numerical aspects of the KMS condition and rather 
general holomorphic extension aspects have recently been studied 
in \cite{dMV16}. \\


{\bf Notation:} We follow the ``physics convention'' that 
the scalar product $\la \cdot,\cdot \ra$ on a complex Hilbert space 
is linear in the second argument. 

For a real vector space $V$, we write $\Bil(V)$ for the complex vector space of 
complex-valued bilinear forms $V \times V \to \C$. 
For $h \in \Bil(V)$, we write $\oline h$ for the pointwise 
complex conjugate  
and put $h^\top(v,w) := h(w,v)$ and $h^* := \oline h^\top$. 
We say that $h$ is {\it hermitian} if $\oline h = h^\top$, which 
means that $\Re h$ is symmetric and $\Im h$ is skew-symmetric. 
We write $\Herm(V) \subeq \Bil(V)$ for the real subspace 
of hermitian forms. 

Every $h \in \Bil(V)$ extends canonically to a sesquilinear form  on $V_\C$ 
(linear in the second argument) 
\[ h_\C(v + i w, v' + i w') := h(v,v') 
- i h(w,v') + i h(v,w') + h(w,w'). \] 
We may therefore identify $\Bil(V)$ with the space 
$\Sesq(V_\C)$ of sesquilinear forms on the complex vector space~$V_\C$.
We write $\Bil^+(V) \subeq \Bil(V)$ for the convex cone of all those 
bilinear forms $f$ for which the sesquilinear extension to $V_\C \times V_\C$ 
is positive semidefinite, i.e., for which $h$ defines a positive definite kernel 
on~$V$. 


\tableofcontents

\section{Positive definite functions and KMS conditions} 

Throughout this section $V$ is an arbitrary real vector space. 
We recall from Definition~\ref{def:a.3} that a function 
$\psi \: \R \to \Bil(V)$ is called {\it positive definite} if the kernel 
$K((t,v), (s,w)) := \psi(t-s)(v,w)$ on $\R \times V$ is positive definite. 
The main result of this section is Theorem~\ref{thm:kms}, 
which is Theorem~\ref{thm:1-intro} from the introduction. This result 
leads in particular to the analytic continuation of $\psi$ to 
the strip $\cS_\beta$. We also explain how the corresponding 
representation of $\R$ can be realized in a Hilbert space consisting 
of holomorphic functions on the strip $\cS_{\beta/2}$ with continuous boundary 
values (Proposition~\ref{prop:2.4}). 

We call a function $\psi \: \oline{\cS_\beta} \to \Bil(V)$ 
{\it pointwise continuous} if, for all $v,w \in V$, the function 
$\psi^{v,w}(z) := \psi(z)(v,w)$ is continuous. 
Moreover, we say that $\psi$ is {\it pointwise holomorphic 
in $\cS_\beta$}, if, for all $v,w \in V$, the 
function $\psi^{v,w}\res_{\cS_\beta}$ is holomorphic. 
By the Schwarz reflection principle, any pointwise continuous 
pointwise holomorphic function $\psi$ is uniquely determined by 
its restriction to~$\R$. 

\begin{defn} \mlabel{def:2.1} 
For $\beta > 0$, let $\cS_\beta := \{ z \in \C \: 0 < \Im z < \beta\}$. 
For a real vector space $V$, we say that a positive definite 
function $\psi \: \R \to \Bil(V)$ 
satisfies the {\it KMS condition} 
for $\beta > 0$ if $\psi$ 
extends to a function $\psi \: \oline{\cS_\beta} \to \Bil(V)$ which is pointwise 
continuous, pointwise holomorphic on $\cS_\beta$, and satisfies 
\begin{equation}
  \label{eq:kms-gen} 
\psi(i \beta+t) = \oline{\psi(t)} 
\quad \mbox{ for } \quad t \in \R. 
\end{equation}
\end{defn} 

\begin{lem}
  \mlabel{lem:2.2} 
Suppose that $\psi \: \R \to \Bil(V)$ satisfies the KMS 
condition for $\beta > 0$. 
Then 
\begin{equation}
  \label{eq:herm-cond}
 \psi(-\oline z) = \psi(z)^* 
\quad \mbox{ and }\quad 
\psi(i\beta + \oline z) = \oline{\psi(z)} 
\quad \mbox{ for }\quad z \in \oline{\cS_\beta}.
\end{equation}
The function $\phi \: [0,\beta] \to \Bil(V), 
\phi(t) := \psi(it)$ has hermitian values and satisfies 
\begin{equation}
  \label{eq:kms-imag} 
\phi(\beta - t) = \oline{\phi(t)}
\quad \mbox{ for } \quad 0 \leq t \leq \beta.
\end{equation}
It extends to a unique pointwise 
continuous symmetric $2\beta$-periodic function $\phi :\R \to \Herm(V)$ 
satisfying 
\[ \phi(\beta + t) = \oline{\phi(t)} \quad \mbox{  for } \quad t \in \R.\] 
\end{lem}

\begin{prf} Note that $\psi(-t) = \psi(t)^*$ 
holds for every positive definite function $\psi \:  \R \to \Bil(V)$.
By analytic continuation (resp., the Schwarz reflection principle), this leads to 
the first part of \eqref{eq:herm-cond}. Likewise, 
condition \eqref{eq:kms-gen} leads  to the second part of 
\eqref{eq:herm-cond}. 
This in turn implies \eqref{eq:kms-imag}, and the remainder is clear.
\end{prf}

\begin{rem} Note that \eqref{eq:herm-cond} implies in particular that 
$\psi(i\beta/2 + t)$ is real-valued for $t \in \R$ 
(cf.\ \cite[Prop.~3.5]{RvD77}). 
\end{rem}

We now introduce standard real subspaces $V \subeq \cH$ 
and the associated modular objects $(\Delta, J)$.  

\begin{defn} \mlabel{def:1.5} 
A closed real subspace $V$ of a complex Hilbert space $\cH$ 
is said to be {\it standard} if 
\[ V \cap i V = \{0\} \quad \mbox{ and } \quad \oline{V + i V} = \cH.\] 

 For every standard real subspace $V \subeq \cH$,  
we define an unbounded antilinear operator 
\[ S \: \cD(S) = V + i V \to \cH, \quad 
S(v + i w) := v - i w, \quad v,w \in V.\] 
Then $S$ is closed and has a polar decomposition 
$S = J \Delta^{1/2},$ 
where $J$ is an anti-unitary involution and $\Delta$ a positive selfadjoint 
operator (cf.\ \cite[Lemma~4.2]{NO15}; see also \cite[Prop.~2.5.11]{BR02}, 
\cite[Prop.~3.3]{Lo08}). 
We call  $(\Delta, J)$ the {\it modular objects} of $V$.
\end{defn}

\begin{rem} \mlabel{rem:2.2b}
(a) From $S^2 = \id$, it follows that 
the modular objects $(\Delta, J)$ of a standard real subspace 
satisfies the modular relation 
\begin{equation}
  \label{eq:modrel}
J\Delta J = \Delta^{-1}.   
\end{equation}
If, conversely, $(\Delta, J)$ is a pair 
of a positive selfadjoint operator $\Delta$ and an antilinear involution $J$ 
satisfying \eqref{eq:modrel}, then $S := J \Delta^{1/2}$ is an unbounded 
antilinear involution with $\cD(S) = \cD(\Delta^{1/2})$ whose fixed point space 
$\Fix(S)$ is a standard real subspace. Thus standard real subspaces  
are parametrized by pairs $(\Delta, J)$ satisfying
\eqref{eq:modrel} (cf.\ \cite[Prop.~3.2]{Lo08} and \cite[Lemma~4.4]{NO15}). 

(b) As the unitary one-parameter group $\Delta^{it}$ commutes with $J$ and $\Delta$, 
it leaves the real subspace $V = \Fix(S)$ invariant. 
\end{rem}

We now come to the proof of Theorem~\ref{thm:1-intro} from the introduction. 
\begin{thm} {\rm(Characterization of the KMS condition)}  \mlabel{thm:kms} 
Let $V$ be a real vector space and 
let $\psi \:  \R \to \Bil(V)$ be a pointwise continuous positive definite function. 
Then the following are equivalent: 
\begin{itemize}
\item[\rm(i)] $\psi$  satisfies the 
KMS condition for $\beta > 0$. 
\item[\rm(ii)] There exists a standard real subspace 
$V_1$ in a Hilbert space $\cH$ and a linear map 
$j \: V \to V_1$ such that 
\begin{equation}
  \label{eq:pdform2}
 \psi(t)(v,w) = \la j(v), \Delta^{-it/\beta}  j(w) \ra \quad \mbox{ for } \quad 
t \in \R,v,w \in V.
\end{equation}
\item[\rm(iii)] There exists a $\Bil^+(V)$-valued regular Borel measure $\mu$ 
on $\R$ satisfying $d\mu(-\lambda) = e^{-\beta\lambda} d\oline\mu(\lambda)$, 
such that 
\[ \psi(t) = \int_\R e^{it\lambda}\, d\mu(\lambda) = \hat\mu(t).\] 
\end{itemize}
If these conditions are satisfied, then the function 
$\psi \: \oline{\cS_\beta} \to \Bil(V)$ is pointwise bounded. 
\end{thm}

\begin{prf} (i) $\Rarrow$ (ii): From the GNS construction 
(Proposition~\ref{prop:gns}) we obtain a continuous unitary representation 
$(U,\cH)$ and a linear map $j \: V \to \cH$ such that 
\[ \psi(t)(v,w) = \la j(v), U_t  j(w) \ra \quad \mbox{ for } \quad 
t \in \R, v,w \in V.\]
We further assume that the range of the map 
\[ \zeta\: \R\times V \to \cH, \quad 
\zeta(t,v) := U_t j(v) \]  
spans a dense subspace. 
Using Stone's Theorem, we write $U_t = e^{-itH}$ for a selfadjoint operator 
$H$ on $\cH$ and consider the positive selfadjoint operator 
\[ \Delta := e^{\beta H} \quad \mbox{ satisfying  } \quad 
U_t = \Delta^{-it/\beta} \quad \mbox{ for } \quad t \in \R.\] 
With the $B(\cH)$-valued spectral measure $P$ on $\R$ with 
$H = \int_\R \lambda\, dP(\lambda)$, we thus obtain 
\[ \psi(t)(v,w) = \la j(v), e^{-itH} j(w) \ra 
= \int_\R e^{-it \lambda}\, dP^{j(v),j(w)}(\lambda),\]
where $P^{v,w} = \la v,P(\cdot )w\ra$. 
The KMS condition for $\psi$ entails that, 
for each $v \in V$, the function 
$\psi(t)(v,v)$ extends holomorphically to $\oline{\cS_\beta}$,  
which entails that the integral 
$\int_\R e^{\beta \lambda}\, dP^{j(v),j(v)}(\lambda)$ is finite, and hence that 
$j(V) \subeq \cD(\Delta^{1/2})$ (\cite[Lemma~B.4]{NO15}). 
The uniqueness of analytic continuation (Schwarz' principle) 
now implies that 
\begin{equation}
  \label{eq:anacont1}
\psi(x + iy)(v,w) 
= \int_\R e^{-i(x+iy) \lambda}\, dP^{j(v),j(w)}(\lambda) 
= \la  \Delta^{y/2\beta}j(v), \Delta^{-ix/\beta}\Delta^{y/2\beta} j(w) \ra 
\end{equation}
for $v,w \in V$ and $0 \leq y \leq \beta$. 
Since $\cD(\Delta^{1/2})$ is $U$-invariant, we obtain  from  the KMS condition 
\begin{align*}
\la \Delta^{1/2} \zeta(t,v), \Delta^{1/2} \zeta(s,w)\ra 
&= \la \Delta^{1/2} j(v), \Delta^{1/2}U_{s-t}j(w)\ra 
= \psi(i\beta + s - t)(v,w) \\
&= \oline{\psi(s - t)(v,w)}
= \oline{\la \zeta(t,v), \zeta(s,w)\ra}.
\end{align*}
This implies the existence of a unique antilinear isometry 
$J\: \cH \to \cH$ with 
\[ J \zeta(t,v) = \Delta^{1/2} \zeta(t,v) \quad\mbox{ for all } \quad 
t \in \R, v \in V.\] 
Then 
\[ U_s J \zeta(t,v) = \Delta^{1/2} \zeta(t+s,v) 
= J \zeta(t+s,v) = J U_s \zeta(t,v) 
\quad \mbox{ for } \quad t,s \in \R, v \in V \] 
shows that $J$ commutes with every $U_t$. This implies that 
$J \Delta^{1/2} J^{-1} = \Delta^{-1/2}$, so that 
\[ \zeta(t,v) = J^{-1} \Delta^{1/2} \zeta(t,v) = \Delta^{-1/2} J^{-1} \zeta(t,v),\] 
which in turn implies 
\[ J\zeta(t,v) = \Delta^{1/2} \zeta(t,v) = J^{-1} \zeta(t,v) 
\quad \mbox{ for } \quad t \in \R, v \in V.\] 
Since the range of $\zeta$ is total, it follows that 
$J^{-1} = J$, so that $J$ is an anti unitary involution. 
Therefore $(\Delta, J)$ is the modular objects of the standard real subspace 
$V_1 := \Fix(S)$ for the unbounded antilinear involution $S := J \Delta^{1/2}$ 
(Remark~\ref{rem:2.2b}). 

For $v \in V$ we now have $j(v) \in \cD(S) = \cD(\Delta^{1/2})$ and 
$Sj(v) = J\Delta^{1/2}j(v) = J^2 j(v) = j(v)$, so that 
$j(V) \subeq V_1$. This completes the proof of (ii). 

(ii) $\Rarrow$ (iii): For $v,w \in V$ we have 
\[ \psi(t)(v,w) = \la j(v),  \Delta^{-it/\beta} j(w) \ra 
= \int_\R e^{it\lambda} \la j(v), dP(\lambda)j(w)\ra,\] 
where $P$ is the spectral measure of the selfadjoint operator 
$L := -\frac{1}{\beta}\log \Delta$ (the Liouvillean). 
We therefore consider the $\Bil^+(V)$-valued measure defined by 
\[ \mu(\cdot)(v,w) := \la j(v),P(\cdot) j(w) \ra = P^{j(v),j(w)}.\] 
It remains to show that 
$d\mu(-\lambda) = e^{-\beta\lambda}d\oline\mu(\lambda)$, 
which means that $r_*\mu = e_{-\beta} \oline\mu$ 
holds for $r(\lambda) = - \lambda$. To verify this relation, 
we first observe that $JLJ = -L$ implies that $JPJ = r_* P$. This leads to 
\begin{align*}
\oline{\mu(\cdot)(v,w)} 
&= \la P(\cdot) j(w) , j(v) \ra =\la P(\cdot) S j(w) , S j(v) \ra 
=\la P(\cdot) J \Delta^{1/2} j(w) , J \Delta^{1/2} j(v) \ra \\
&=\la JP(\cdot) J \Delta^{1/2} j(v) , \Delta^{1/2} j(w) \ra 
=\la (r_*P)(\cdot) \Delta^{1/2} j(v) , \Delta^{1/2} j(w) \ra \\
&=e_{\beta} \cdot  \la (r_*P)(\cdot) j(v) , j(w) \ra 
= e_{\beta}\cdot (r_*\mu)(\cdot)(v,w). 
\end{align*}
This implies that 
$\oline\mu = e_{\beta} \cdot r_*\mu.$

(iii) $\Rarrow$ (i): Condition (iii) implies that 
$\psi(0) = \mu(\R)$ exists, so that $\mu$ is a pointwise finite measure. 
Further, the relation $r_*\mu = e_{-\beta} \oline\mu$ implies that 
the measure $e_{-\beta} \mu$ is also finite. Therefore the integral 
\begin{equation}
  \label{eq:ftrafo}
 \psi(z) := \int_\R e^{iz\lambda}\, d\mu(\lambda) 
\end{equation}
exists pointwise and 
extends $\psi$ to $\oline{\cS_\beta}$ in such a way that this extension 
is pointwise continuous on $\oline{\cS_\beta}$ and pointwise holomorphic on the interior. 
The relation $r_* \mu = e_{-\beta} \oline \mu$ further leads to 
\begin{align*}
\psi(i\beta + t) 
&= \int_\R e^{\lambda(-\beta + it)}\, d\mu(\lambda)
= \int_\R e_{-\beta}(\lambda) e^{i\lambda t}\, d\mu(\lambda)
= \int_\R  e^{i\lambda t}\, d(r_*\oline\mu)(\lambda)
= \int_\R  e^{-i\lambda t}\, d\oline\mu(\lambda)
= \oline{\psi(t)}. 
\end{align*}
Therefore $\psi$ satisfies the KMS condition for $\beta$. 

We finally assume that (i)-(iii) are satisfied 
and show that $\psi$ is pointwise bounded on $\oline{\cS_\beta}$. 
Since each $\psi(z)$ extends to a sesquilinear form $\psi(z)_\C$ 
on $V_\C$, in view of the 
Polarization Identity, it suffices to show the 
boundedness of the functions $z \mapsto \psi(z)_\C(v,v)$ for $v \in V_\C$. 
For the positive measure $\mu^{v,v}(E) := \mu(E)_\C(v,v)$, we obtain from 
\eqref{eq:ftrafo} the estimate 
\[ |\psi(z)_\C(v,v)|
\leq \int_\R |e^{-i\lambda z}|\, d\mu^{v,v}(\lambda) 
= \int_\R e^{\lambda\Im z}\, d\mu^{v,v}(\lambda). \] 
The convexity of the function on the right, the Laplace transform 
of the finite positive measure $\mu^{v,v}$, and 
$\psi(\beta i)(v,v) = \|\Delta^{1/2}j(v)\|^2 < \infty$ now imply the 
boundedness of $\psi(z)_\C(v,v)$. 
\end{prf}

\begin{rem} An important special case arise 
from a $C^*$-dynamical system $(\cA,\R,\alpha)$ 
for $V := \cA_h := \{A  \in \cA \:  A^* = A\}$ and 
an invariant state $\omega$ on $\cA$. 
Such a state is a $\beta$-KMS state if and only if 
\[ \psi \: \R \to \Bil(\cA_h), \quad 
\psi(t)(A,B) :=\omega(A\alpha_t(B)) \] 
satisfies the KMS condition for $\beta > 0$  
(cf.\  \cite[Prop.~5.2]{NO15}, \cite[Thm.~4.10]{RvD77}, \cite{BR96}). 
If $(\pi_\omega, U^\omega, \cH_\omega, \Omega)$ 
is the corresponding covariant GNS representation of $(\cA,\R)$, 
then 
\[ \omega(A) = \la \Omega, \pi_\omega(A) \Omega \ra \quad \mbox{ for }\quad 
A \in \cA \quad \mbox{ and }\quad 
U^\omega_t \Omega = \Omega \quad \mbox{ for }\quad t \in \R.\] 
Therefore 
\[ \psi(t)(A,B) = \omega(A\alpha_t(B)) 
= \la \Omega, \pi_\omega(A\alpha_t(B)) \Omega \ra 
= \la \Omega, \pi_\omega(A) U^\omega_t \pi_\omega(B) U^\omega_{-t} \Omega \ra 
= \la \pi_\omega(A)\Omega,  U^\omega_t \pi_\omega(B) \Omega \ra \] 
for $A, B \in \cA_h$. The corresponding standard real subspace of 
$\cH_\omega$ is $V_1 := \oline{\pi_\omega(\cA_h)\Omega}$. 
\end{rem}

\begin{cor}
  \mlabel{cor:5.2} If $\psi \:  \R \to \Bil(V)$ satisfies
the $\beta$-KMS condition, then the kernel 
\begin{equation}
  \label{eq:kernel}
 K \: \oline{\cS_{\beta/2}} \times \oline{\cS_{\beta/2}} \to \Bil(V),\qquad 
K(z,w)(\xi,\eta) := \psi(z-\oline w)(\xi,\eta)
\end{equation}
is positive definite. 
\end{cor}

\begin{prf} This follows immediately from the following relation 
that we derive from \eqref{eq:anacont1}: 
\[ K(z,w)(\xi,\eta) = \psi(z-\oline w)(\xi,\eta) 
= \la \Delta^{-\frac{i \oline z}{\beta}} j(\xi), \Delta^{-\frac{i\oline w}{\beta}} j(\eta) \ra 
\quad \mbox{ for } \quad \xi,\eta\in V,\ z,w \in \oline{\cS_{\beta/2}}.
\qedhere\] 
\end{prf}

Now that we know from Corollary~\ref{cor:5.2} that the kernel 
$K$ in \eqref{eq:kernel} 
is positive definite, we obtain a corresponding reproducing kernel 
Hilbert space consisting of functions on $\oline{\cS_{\beta/2}} \times V$ 
which are linear in the second argument and holomorphic on 
$\cS_{\beta/2}$ in the first. We may therefore think of these 
functions as having values in the algebraic dual space 
$V^* := \Hom(V,\R)$ of~$V$. 
We write 
$\cO(\oline{\cS_{\beta/2}},V^*)$ for the space of those functions 
$f \: \oline{\cS_{\beta/2}}\to V^*$ with the property that, for every 
$\eta \in V$, the function $z \mapsto f(z)(\eta)$ is continuous on 
$\oline{\cS_{\beta/2}}$ and holomorphic on the open strip~$\cS_{\beta/2}$. 

\begin{prop} \mlabel{prop:2.4} 
{\rm(Realization of $\cH_\psi$ on $\cO(\oline{\cS_{\beta/2}},V^*)$)} 
Assume that $\psi \: \R \to \Bil(V)$ satisfies the KMS 
condition for $\beta > 0$ and let $\psi \: \oline{\cS_\beta} \to \Bil(V)$ 
denote the corresponding extension and 
$\cH_\psi \subeq \cO(\oline{\cS_{\beta/2}},V^*)$ denote the Hilbert space with reproducing 
kernel 
\[ K(z,w)(\xi,\eta)  := \psi(z- \oline w)(\xi,\eta) 
\quad \mbox{ for } \quad \xi,\eta \in V,\] 
i.e., 
\[ f(z)(\xi) = \la K_{z,\xi}, f \ra \quad \mbox{ for } \quad f \in \cH_\psi, 
\quad \mbox{ where } \quad 
K_{z,\xi}(w)(\eta) = \psi(w - \oline z)(\eta,\xi).\] 
Then 
\[ (U^\psi_t f)(z) := f(z + t), \qquad t \in \R, z \in \oline{\cS_{\beta/2}}\]
defines a unitary one-parameter group on $\cH_\psi$, 
\[ j \: V \to \cH_\psi, \quad j(\eta)(z) := \psi(z)(\cdot, \eta) 
\] 
is a linear map with $U^\psi$-cyclic range, and 
\[ \psi(t)(\xi,\eta) = \la j(\xi), U^\psi_t j(\eta) \ra \quad \mbox{ for } \quad 
t\in \R, \xi,\eta \in V.\] 
The anti-unitary involution on $\cH_\psi$ corresponding to 
the standard real subspace $V_1 \subeq \cH_\psi$ from 
{\rm Theorem~\ref{thm:kms}} is given by 
\begin{equation}
  \label{eq:J-repro}
(J_1 f)(z) := \oline{f\Big(\oline z + \frac{i\beta}{2}\Big)}. 
\end{equation}
\end{prop}

\begin{prf} First we recall that the natural 
reproducing kernel Hilbert space $\cH_\psi = \cH_K$ is 
generated by function $K_{(w,\eta)}$ satisfying 
\[ K_{(w,\eta)}(z)(\xi) = \la K_{(z,\xi)}, K_{(w,\eta)} \ra 
= K(z,w)(\xi,\eta)= \psi(z - \oline{w})(\xi,\eta). \] 
As a function of $z$, the kernel $K$ is continuous 
on $\oline{\cS_{\beta/2}}$ and holomorphic on the interior. 
Therefore \cite[Prop.~I.1.9]{Ne00} implies that 
$\cH_\psi$ is a subspace of $\cO(\oline{\cS_{\beta/2}},V^*)$, 
where, for every $f \in \cH_\psi$ and $\xi \in V$, we have 
\[ f(z)(\xi) = \la K_{(z,\xi)}, f \ra. \] 
That the formula for $U^\psi_t$ defines a unitary one-parameter 
group on $\cH_\psi$ follows directly from the invariance of the kernel~$K$ 
under the action of $\R$ on $\oline{\cS_\beta}$ by translation.  

Next we observe that 
\[ \la j(\xi), U^\psi_t j(\eta) \ra
= \la K_{(0,\xi)}, U^\psi_t K_{(0,\eta)}\ra
= \la K_{(0,\xi)}, K_{(-t,\eta)}\ra
= \psi(t)(\xi,\eta).\] 

To see that $j(V)$ is $U^\psi$-cyclic, we have to show that 
the elements $U^\psi_t j(\eta) = K_{(-t,\eta)}$ form a total subset of $\cH_\psi$. 
This means that any $f \in \cH_\psi$ with 
$0 = \la K_{(t,\eta)}, f \ra = f(t)(\eta)$ 
for every $t \in \R$ and $\eta \in V$ vanishes. 
As the function $t \mapsto f(t)(\eta)$ extends to a continuous 
function on $\oline{\cS_{\beta/2}}$, holomorphic on the interior, 
it vanishes by the Schwarz Reflection Principle. Further, $\eta$ was arbitrary,
so that $f = 0$ follows. 

Now we turn to the involution $J_1$. 
As $K_{(w,\eta)}(z) = \psi(z- \oline w)(\cdot,\eta)$, the operator 
$J_1$ on $\cO(\oline{\cS_{\beta/2}}, V^*)$, defined by the right hand side of 
\eqref{eq:J-repro} satisfies 
\begin{eqnarray} 
(J_1 K_{(w,\eta)})(z) 
&=& \oline{K_{(w,\eta)}\Big(\oline z + \frac{i\beta}{2}\Big)}
= \oline{\psi\Big(\oline z + \frac{i\beta}{2} - \oline w\Big)(\cdot,\eta)}
= \psi\Big(i\beta + z - \frac{i\beta}{2} - w\Big)(\cdot,\eta)\notag\\
&=& \psi\Big(z + \frac{i\beta}{2} - w\Big)(\cdot,\eta)
= K_{(\oline w  + i\beta/2, \eta)}(z). 
\end{eqnarray}
Here we have used that $\oline{\psi(z)} = \psi(i\beta + \oline z)$ 
(Lemma~\ref{lem:2.2}). 
>From 
\begin{align*}
\la K_{(\oline w  + i\beta/2, \eta)}, K_{(\oline z  + i\beta/2, \xi)} \ra 
&= \oline{K(\oline z + i \beta/2, \oline w + i \beta/2)(\xi,\eta)} 
= \oline{\psi(i\beta + \oline z - w)(\xi,\eta)} \\
&= \psi(z - \oline w)(\xi,\eta) 
= \la K_{(z,\xi)}, K_{(w,\eta)} \ra
\end{align*}
it now follows that the operator $J_1$ in \eqref{eq:J-repro} leaves 
the subspace $\cH_\psi$ invariant and defines an antilinear 
isometry on this space. From the explicit formula it follows that 
$J_1$ is an involution. It is also clear that 
$J_1$ commutes with the unitary operators 
$(U_t f)(z) = f(z + t)$. 

The relation $U_t K_{(w,\eta)} = K_{(w-t,\eta)}$ leads by analytic continuation to 
\[ J_1 K_{(0,\eta)} = K_{(i\beta/2, \eta)} = \Delta^{1/2} K_{(0,\eta)}.\]
In the proof of Theorem~\ref{thm:kms}, we have seen that, for $\eta \in V$ and 
$t\in \R$, the anti-unitary involution $J$ corresponding to the associated standard 
real subspace $V_1$ satisfies 
\[ J j(\eta) = \Delta^{1/2} j(\eta).\] 
As both, $J$ and $J_1$ commute with every $U_t$ and 
the subset $\{U_t j(\eta) \: t \in \R, \eta \in V \}$ is total in 
$\cH_\psi$, we conclude that $J_1 = J$. 
\end{prf}


\section{Form-valued reflection positive functions} 

In this section we discuss reflection positivity on the level of 
form-valued positive definite function. This is particularly well-adapted 
to reflection positive Hilbert spaces  
$(\cE,\cE_+,\theta)$, for which 
$\cE_+$ is generated by elements of the form 
$U_g^{-1} j(v)$, where $g$ is contained in a certain subset $G_+ \subeq G$ 
which is not necessarily a subsemigroup, and $j\: V \to \cH$ is a linear 
map for which $U_G j(V)$ spans a dense subspace of $\cE$. 
In particular, we present a version of the GNS construction in this 
context (Proposition~\ref{prop:1.x}) and we briefly discuss it more specifically 
for the trivial group $G = \{\1\}$ (Subsection~\ref{subsec:3.2}) 
and the $2$-element group (Subsection~\ref{subsec:3.3}). 
The latter case shows explicitly that the cone of reflection positive 
functions does not adapt naturally to the decomposition 
into even and odd functions. Put differently, if a reflection positive 
representation decomposes into two subrepresentations, the summands 
need not be reflection positive (see also \cite{NO14a}).

\subsection{Reflection positivity and form-valued functions} 
\mlabel{subsec:3.1}

Let $(G,\tau)$ be a {\it symmetric Lie group}, i.e., $G$ is a Lie group and 
$\tau \in \Aut(G)$ with $\tau^2 = \id_G$. 
In the following we write $G_\tau := G \rtimes \{\1,\tau\}$ 
and $g^\sharp :=\tau(g)^{-1}$ (\cite{NO14a}). 
In this section we introduce reflection positive 
functions on $G_\tau$ with values in $\Bil(V)$ for a real vector space~$V$. 

\begin{defn} \mlabel{def:x.1} Let $\cE$ be a Hilbert space and 
let $\theta \in \U(\cE)$ be an involution. 
A closed subspace $\cE_+ \subeq \cE$ is called {\it $\theta$-positive} 
if $\la \theta v,v \ra \geq 0$ for $v \in \cE_+$. 
We then call the triple $(\cE,\cE_+,\theta)$ a {\it reflection positive 
Hilbert space}. For a reflection positive Hilbert space we put 
$\cN := \{ v \in \cE_+ \: \la \theta v, v \ra = 0\}$ 
and write \break $q \: \cE_+ \to \cE_+/\cN, v \mapsto \hat v = q(v)$ for the quotient map and $\hat\cE$ for the Hilbert completion of $\cE_+/\cN$ with respect to 
the norm $\|\hat v\|_{\hat{\cE}}:=\|\hat v\| := \sqrt{\la \theta v, v \ra}$. 
\end{defn}

\begin{ex} \mlabel{ex:3.2} 
Suppose that $K \colon X \times X \to \C$ is a positive definite kernel 
on the set $X$ and that $\tau \colon X \to X$ is an involution 
leaving $K$ invariant. We further assume that 
$X_+ \subeq X$ is a subset with the property that the 
kernel $K^\tau(x,y) := K(\tau x, y)$ is also positive definite on $X_+$. 

Let $\cE := \cH_K\subeq \C^X$ denote the corresponding reproducing kernel 
Hilbert space generated by elements $(K_x)_{x \in X}$ 
with $\la K_x, K_y \ra = K(x,y)$. 
Then the closed subspace $\cE_+ \subeq \cE$ generated 
by $(K_x)_{x \in X_+}$ is $\theta$-positive for 
$(\theta f)(x) := f(\tau x)$. We thus obtain a 
reflection positive Hilbert space $(\cE,\cE_+,\theta)$. 
We call such kernels $K$ {\it reflection positive} with respect to $(X,X_+, \tau)$. 
\end{ex} 

\begin{defn}
  \mlabel{def:1.2} Let $G_+ \subeq G$ be a subset. 
Let $V$ be a real vector space and let $j \: V \to \cH$ be a linear map whose 
range is cyclic for the unitary representation $(U, \cE)$ of $G_\tau$. 
Then we say that $(U, \cE, j,V)$ is 
{\it reflection positive with respect to $G_+ \subeq G$} if, for 
 $\cE_+ := \oline{\Spann U_{G_+}^{-1} j(V)}$, 
the triple  $(\cE,\cE_+,U_\tau)$ is a reflection positive Hilbert space. 
\end{defn}

\begin{defn}\mlabel{def:1.2c} 
Let $V$ be a real vector space. We call a function 
$\phi: G_\tau  \to \Bil(V)$ {\it reflection positive with respect to 
the subset $G_+$ of $G$} if 
\begin{description}
\item[\rm(RP1)] $\phi$ is positive definite and  
\item[\rm(RP2)] the kernel $(s,t) \mapsto 
\phi(s t^\sharp \tau) = \phi(s\tau t^{-1})$ is positive definite on~$G_+$. 
\end{description}
\end{defn} 

\begin{rem} Let $\phi \: G_\tau \to \Bil(V)$ be a positive definite 
function, so that the kernel 
$K((x,v), (y,w)) := \phi(xy^{-1})(v,w)$ on $G_\tau \times V$ is positive definite. 
The involution $\tau$ acts on $G_\tau \times V$ by 
$\tau.(g,v) := (g\tau,v)$ and the corresponding kernel 
$K^\tau((x,v),(y,w)) := K((x\tau,v),(y,w)) 
= \phi(x\tau y^{-1})(v,w)$ is positive definite on 
$G_+ \times V$ if and only if $\phi$ is reflection positive 
in the sense of Example~\ref{ex:3.2}. 
\end{rem}

Positive definite functions  on $G$ extend canonically to 
$G_\tau$ if they are $\tau$-invariant: 

\begin{lem} \mlabel{lem:biinvar} Let $V$ be a real vector space 
and let $(G,\tau)$ be a symmetric Lie group. Then the following assertions hold: 
\begin{itemize}
\item[\rm(i)] 
If $\phi\: G \to \Bil(V)$ is a positive definite function 
which is $\tau$-invariant in the sense that $\phi \circ \tau = \phi$, then 
$\hat\phi(g,\tau)  := \phi(g)$ defines an extension to 
$G_\tau$ which is positive definite and $\tau$-biinvariant. 
\item[\rm(ii)] Let $(U,\cH)$ be a unitary 
representation of $G_\tau$, let $\theta := U_\tau$, let $j \: V \to \cH$ be a linear map, 
 and let $\phi(g)(v,w) = \la j(v), U_g  j(w)\ra$ be the corresponding $\Bil(V)$-valued 
positive definite function. Then the following are equivalent: 
\begin{itemize}
\item[\rm(a)] $\theta j(v) = j(v)$ for every $v \in V$. 
\item[\rm(b)] $\phi$ is $\tau$-biinvariant.  
\item[\rm(c)] $\phi$ is left $\tau$-invariant.  
\end{itemize}
\end{itemize}
\end{lem}

\begin{prf} (i) From the GNS construction 
(Proposition~\ref{prop:gns}), we obtain a continuous unitary representation 
$(U,\cH)$ of $G$ and a linear map $j \: V \to \cH$ such that 
\[ \phi(g)(v,w) = \la j(v), U_g j(w) \ra \quad \mbox{ for } \quad g \in G, v,w \in V.\]
As $\phi(g)(v,w) = \phi(\tau(g))(v,w)$, the uniqueness in the GNS construction 
provide a unitary operator \break $\theta \: \cH \to \cH$ with 
\[ \theta U_g j(v) = U_{\tau(g)} j(v) \quad \mbox{ for } \quad g \in G, v \in V.\] 
Note that $\theta$ fixes each $j(v)$. 
Therefore $U_{\tau} := \theta$ defines an extension of $G$ to a unitary 
representation of $G_\tau$ on $\cH$. Hence 
$\psi(g)(v,w) = \la j(v), U_g j(w) \ra$ defines a positive definite 
$\Bil(V)$-valued function on $G_\tau$ which satisfies 
\[ \psi(g,\tau)(v,w) 
= \la \theta j(v), U_g j(w) \ra 
= \la j(v), U_g j(w) \ra 
= \phi(g)(v,w) \quad \mbox{ for } \quad g \in G,v,w \in V.\] 

(ii) Clearly, (a) $\Rarrow$ (b) $\Rarrow$ (c). 
It remains to show that (c) implies (a). So we assume that 
$\phi(\tau g) = \phi(g)$ for $g \in G_\tau$. This means that, for every 
$v,w \in V$, we have
\[ \la j(v), U_g j(w) \ra = \phi(g)(v,w) 
= \phi(\tau g)(v,w) = \la j(v), \theta U_g j(w) \ra 
= \la \theta j(v), U_g j(w) \ra.\] 
Since $U_{G_\tau}j(V)$ is total in $\cH$, this implies that 
$\theta j(v) = j(v)$ for every $v \in V$. 
\end{prf}

\begin{rem} \mlabel{rem:1.3} 
(a) As $G_\tau$ consists of the two cosets $G$ and $G\tau  
= G \times \{\tau\}$, every function $\phi$ on $G_\tau$ is given by a 
pair of functions on $G$: 
\[ \phi_\pm \: G \to \Bil(V), \quad 
\phi_+(g) := \phi(g,\1), \qquad \phi_-(g) := \phi(g,\tau).\] 
Then (RP2) is a condition on $\phi_-$ alone, and (RP1) is a condition 
on the pair $(\phi_+, \phi_-)$. 

(b) If $\phi$ is reflection positive, then 
its complex conjugate $\oline\phi$ is also reflection positive 
because the convex cone of positive definite kernels on a set 
is stable under complex conjugation. This implies in particular that 
$\Re \phi = \frac{1}{2}(\phi  + \oline \phi)$
is reflection positive (cf.\ Theorem~\ref{thm:1.1}). 
\end{rem}

The following lemma provides a tool which is sometimes convenient to 
verify positive definiteness of a function on the extended  group $G_\tau$ 
in terms of a kernel on the original group~$G$. 

\begin{lem} \mlabel{lem:3.7} 
 Every function $\phi \: G_\tau \to B(V)$ leads to a $M_2(B(V))$-valued 
kernel 
\[ Q \: G \times G \to M_2(B(V))\cong B(V \oplus V), \quad 
Q(g,h) = \pmat{ 
\phi(gh^{-1}) & \phi(g\tau h^{-1})\\ 
\phi(g\tau h^{-1}) & \phi(gh^{-1})},\] 
and the 
function $\phi$ on $G_\tau$ is positive definite if and only if $Q$ 
is positive definite. 
\end{lem}

\begin{prf} That $Q$ is positive definite 
is equivalent to the exists of a Hilbert space $\cH$ and a map 
\[ \ell \: G \to B(\cH,V \oplus V)\cong B(\cH,V)^{\oplus 2} 
\quad \mbox{  with } \quad 
Q(x,y) = \ell(x) \ell(y)^* \quad \mbox{ for } \quad x,y \in G \] 
(cf.\ \cite[Thm.~I.1.4]{Ne00}). 
If $\ell$ is such a map, then it can be written 
as $\ell(x) = (\ell_1(x), \ell_2(x))$ with $\ell_j(x) \in B(\cH,V)$. 
We thus obtain 
\[ Q(x,y)= \ell(x) \ell(y)^* 
= \pmat{ 
\ell_1(x) \ell_1(y)^* & \ell_1(x) \ell_2(y)^* \\ 
\ell_2(x) \ell_1(y)^* & \ell_2(x) \ell_2(y)^*}\] 
and thus
\[ \ell_1(x) \ell_1(y)^*  = \ell_2(x) \ell_2(y)^* \quad \mbox{ and } \quad  
\ell_1(x) \ell_2(y)^*  = \ell_2(x) \ell_1(y)^*.\] 
Therefore 
\[ j \: G_\tau \to B(\cH,V), \quad 
j(x,\1) := \ell_1(x), \qquad j(x,\tau) := \ell_2(x) \] 
satisfies 
\[ j(x,\1) j(y,\1)^* = \ell_1(x) \ell_1(y)^* = \phi(xy^{-1}), \quad  
j(x,\tau) j(y,\tau)^* = \ell_2(x) \ell_2(y)^* = \phi(xy^{-1}) \] 
and 
\[ j(x,\1) j(y,\tau)^* = \ell_1(x) \ell_2(y)^* = \phi(x\tau y^{-1}), \quad  
j(x,\tau) j(y,\1)^* = \ell_2(x) \ell_1(y)^* = \phi(xy^{-1}). \] 
We therefore have 
$\phi(xy^{-1}) = j(x)j(y)^*$ for $x,y \in G_\tau$, and thus $\phi$ 
is positive definite. 

If, conversely, $\phi$ is positive definite and $j \: G_\tau \to B(\cH,V)$ is such that 
$\phi(x^{-1}y) = j(x)j(y)^*$ for $x,y \in G_\tau$, then 
$\ell(x) := (j(x,\1), j(x,\tau)) \in B(\cH,V \oplus V)$ 
defines a map with $Q(x,y) = \ell(x) \ell(y)^*$ for $x,y \in G$. 
\end{prf}

\begin{prop}
  \mlabel{prop:1.x} {\rm(GNS construction for reflection positive functions)} 
Let $V$ be a real vector space, let $(U, \cE)$ be a unitary representation
 of $G_\tau$ and put $\theta := U_{\tau}$. Then the following assertions hold: 
\begin{itemize}
\item[\rm(i)] If $(U, \cH,j,V)$ is  reflection positive w.r.t.\ $G_+$, 
then 
\[ \phi(g)(v,w) := \la j(v), U_g j(w)\ra, \qquad g \in G_\tau, v, w \in V, \]
is a reflection positive $\Bil(V)$-valued function. 
\item[\rm(ii)] If $\phi \: G_\tau \to \Bil(V)$ is a reflection positive function 
w.r.t.\ $G_+$, 
then the corresponding GNS representation 
$(U^\phi, \cH_\phi,j,V)$ is a reflection positive 
representation, where $\cH_\phi \subeq \C^{G_\tau \times V}$ is the Hilbert subspace 
with reproducing kernel 
$K((x,v), (y,w)) := \phi(xy^{-1})(v,w)$ on which $G_\tau$ acts by 
\[ (U^\phi_g f)(x,v) := f(xg,v).\] 
\end{itemize}
\end{prop}

\begin{prf} (i) For $s, t \in G_+$,  we have 
  \begin{align*}
 \phi(s\tau t^{-1})(v,w) 
&= \la j(v), U_{s \tau t^{-1}} j(w)\ra 
= \la U_{s^{-1}} j(v), U_{\tau} U_{t^{-1}} j(w)\ra 
= \la \theta U_{s^{-1}} j(v), U_{t^{-1}} j(w)\ra, 
  \end{align*}
so that the kernel $(\phi(s\tau t^{-1}))_{s,t \in G_+}$ is positive definite. 

(ii) Recall the relation 
$\phi(g)(v,w) = \la j(v), U_g j(w) \ra$ for 
$g \in G, v,w \in V$ from Proposition~\ref{prop:gns}. 
Moreover, $(\theta f)(x,v) = f(x\tau,v)$, and 
\[  \la \theta  U^\phi_{s^{-1}} j(v), U^\phi_{t^{-1}} j(w)\ra 
=  \la j(v), U^\phi_{s \tau t^{-1}} j(w) \ra 
= \phi(s\tau t^{-1})(v,w),\] 
so that the positive definiteness of the kernel 
$(\phi(s\tau t^{-1}))_{s,t \in G_+}$ implies that we obtain with $\cE = \cH_\phi$ and 
$\cE_+ := \oline{\Spann{(U^\phi_{G_+})^{-1}j(V)}}$ a reflection 
positive Hilbert space $(\cE,\cE_+,\theta)$.
\end{prf}

\subsection{Reflection positivity for the trivial group} 
\mlabel{subsec:3.2}

In this short section we discuss the case of the $2$-element 
group $T = \{\1,\tau\}$ in some detail. 
It corresponds to $G_\tau$ where $G= \{\1\}$ is trivial, 
but it already demonstrates how the intricate structure 
of a reflection positive Hilbert space $(\cE,\cE_+,\theta)$ can be encoded 
in terms of positive definite functions on~$T$.

A unitary representation $(U,\cE)$ of $T$ is nothing but the specification 
of a unitary operator $\theta = U_\tau$ on $\cE$. We write 
$\cE = \cE^1 \oplus \cE^{-1}$ for the eigenspace decomposition 
of $\cE$ under~$\theta$ and $p^{\pm 1} \: \cE \to \cE^{\pm 1}$ for the 
orthogonal projections. 

Suppose, in addition, that $V$ be a real or complex Hilbert space and 
that $j \: V \to \cE$ is a continuous linear map whose range generates $\cE$ under the 
representation $U$, i.e., 
the projections $p^{\pm 1}(j(V)) \subeq \cE^{\pm 1}$ are dense subspaces. 
In view of the GNS construction, the data $(\cE,U,j,V)$ is encoded in the 
operator-valued positive definite function 
\[ \phi \: T \to B(V), \quad \phi(g) = j^* U_g j.\] 

For a function $\phi \: T \to B(V)$, let  $A := \phi(\1)$ and $B := \phi(\tau)$. 
Then $\phi$ is positive definite if and only if 
$A = A^* \geq 0$, $B = B^*$, and the operator matrix 
\[ \pmat{ \phi(\1) & \phi(\tau) \\ \phi(\tau) & \phi(\1)} 
= \pmat{ A& B \\ B & A} \in M_2(B(V)) \cong B(V \oplus V)\] 
defines a positive operator (Lemma~\ref{lem:3.7} and \cite[Rem.~I.1.3]{Ne00}). 
This is equivalent to 
\begin{equation}
  \label{eq:AB-esti}
|\la Bv,w \ra|^2 \leq \la Av, v\ra \la Aw,w\ra \quad \mbox{ for } \quad 
v,w \in V
\end{equation}
(cf.\ Corollary~\ref{cor:1.3}). 
Note that \eqref{eq:AB-esti} holds if $A = \1$ and $\|B\| \leq 1$. 
If, more generally, $A$ is invertible, then \eqref{eq:AB-esti} is equivalent to 
$\|A^{-1/2} B A^{-1/2}\| \leq \1.$ 
Here $A= j^*j$ basically encodes how $V$ is mapped into $\cE$ and 
$B$ encodes the unitary involution $\theta$. 

The function $\phi$ is reflection positive w.r.t.\ $G_+ = \{\1\}$ if and only 
if $B = \phi(\tau) \geq 0$, which means that $j(V)$ is $\theta$-positive. 
In this sense reflection positive functions on $T$ encode 
reflection positive Hilbert spaces $(\cE,\cE_+, \theta)$ 
by $\theta = U_\tau$ and $\cE_+ := \oline{j(V)}$.
A pair $(A,B)$ of hermitian operators on $V$ 
corresponds to a reflection positive function $\phi \: T \to B(V)$ 
if and only if $0 \leq B \leq A$. By the Cauchy--Schwarz inequality, 
this is equivalent to \eqref{eq:AB-esti} if $A$ and $B$ are positive operators. This shows that 
\[ \phi= \phi_0 + \phi_1 \quad \mbox{ with } \quad 
\phi_0(\1) = A-B, \quad \phi_0(\tau) = 0 \quad \mbox{ and } \quad 
\phi_1(\1) = \phi_1(\tau) = B,\] 
where both functions $\phi_0$ and $\phi_1$ are reflection positive. 
The function $\phi_0$ corresponds to the case where 
$\cE_+ \bot \theta\cE_+$, so that $\hat\cE = \{0\}$, and 
the constant function $\phi_1$ corresponds to the trivial representation 
of $T$, hence to $\theta = \1$, which means that $q \: \cE_+ \to \hat\cE$ 
is isometric.

Replacing 
$V$ by $\cE_+$, we see that reflection positive functions 
$\phi \: T \to B(\cE_+)$ with $\phi(\1) = \1$ encode reflection 
positive Hilbert spaces $(\cE,\cE_+,\theta)$ for which 
$p^{\pm 1}(\cE_+)$ is dense in $\cE^{\pm 1}$. By \eqref{eq:AB-esti}, 
these configurations are parametrized by the hermitian contractions 
$B = \phi(\tau)$ 
on~$\cE_+$. For $v,w \in \cE_+$, we then have 
\[ \la v, \theta w \ra = \la v, Bw \ra.\]
Therefore the $1$-eigenspace $\ker(B-\1)$ corresponds to the maximal 
subspace in $\cE_+$ on which the map $q \: \cE_+ \to \hat\cE$ is isometric. 
We also observe that $\ker B = \ker q$. In this sense the operator $B$ 
describes how $\hat\cE$ is obtained from the Hilbert space~$\cE_+$.

\begin{rem}
Suppose that $\theta$ is a unitary involution on $\cE$ with the 
eigenspaces $\cE^{\pm 1}$. 
If $\cK \subeq \cE$ is a $\theta$-positive subspace, then clearly 
$\cK \cap \cE^{-1}= \{0\}$ and this implies that 
$\cK$ is the graph $\Gamma(Z)$ of the operator 
\[  Z \: {\cal D}(Z)  := \{ v_+ \in \cE^1 \: (\exists v_- \in \cE^{-1})\ (v_+, v_-) 
\in \cK\} \to \cE^{-1}, \quad v_+ \mapsto v_-. \] 
That $\Gamma(Z)$ is a $\theta $-positive subspace is equivalent to 
$\|Z\| \leq 1$. Therefore the closedness of $\cK$ shows that 
${\cal D}(Z)$ is a closed subspace of $\cE^1$ (cf.\ \cite[Lemma 5.1]{Jo02}). 
If $p^1(\cK) = \cD(Z)$ is dense in $\hat\cE$, the closedness of $\cD(Z)$ implies that 
$Z \in B(\cE^1, \cE^{-1})$. The density of 
$p^{-1}(\cK) = Z(\cE^1)$ is equivalent to $Z$ having dense range. 

>From this perspective, we can also generate the configuration 
$(\cE,\cE_+,\theta)$ in terms of $\cE^1$. Then 
$j(v) = (v,Zv) \in \cE^1 \oplus \cE^{-1}$ defines a linear 
map $j \: \cE^1 \to \cE$ whose range is $\cK$. The corresponding 
$B(\cE^1)$-valued positive definite function on $T$ is given by 
\[ \psi(\1) = j^* j = \1 + Z^* Z \quad \mbox{ and } \quad 
\psi(\tau) = j^*\theta j = \1 - Z^*Z.\] 
The polar decomposition of $j \: \cE^1 \to \cK$ takes the form 
\[ j = U \sqrt{j^*j} = U \sqrt{\1 + Z^*Z},\] 
where $U \: \cE^1 \to \cK$ is unitary. Therefore the corresponding 
$B(\cK)$-valued positive definite function on $T$ is given by 
\[ \phi(\1) = \1 \quad \mbox{ and } \quad 
\phi(\tau) = U \frac{\1 - Z^* Z}{\1 + Z^*Z} U^{-1}\] 
because 
$j^*\phi(\tau)j = j^*\theta j = \1- Z^* Z$ implies 
\[ \phi(\tau) 
= (j^*)^{-1}(\1 - Z^* Z) j^{-1} 
= U (\1 + Z^*Z)^{-1/2}(\1 - Z^* Z) (\1 + Z^*Z)^{-1/2} U^{-1} 
= U \frac{\1 - Z^* Z}{\1 + Z^*Z} U^{-1}.\] 
Relating this to the preceding discussion, 
we see that $U\ker Z \subeq \cE_+$ is the 
maximal subspace on which $q$ is isometric and 
\[ U\{ v \in \cE^1 \: \|Zv\| = \|v\| \} = U \ker(\1 - Z^*Z) = \ker q.\] 
In particular, $q$ is injective if and only if $Z$ is a strict contraction. 
\end{rem}

\subsection{Reflection positivity for the $2$-element group}
\mlabel{subsec:3.3}

In this subsection we take a closer look at the $2$-element group 
$G = \{\1,\sigma\}$ because it nicely illustrates that, 
if a reflection positive 
representation decomposes into two subrepresentations, then the summands 
need not be reflection positive (see also \cite{NO14a}). 
On the level of positive definite functions, this is reflected in the fact that 
the cone of reflection positive functions does not adapt to the decomposition 
into even and odd functions. 

We consider the $2$-element group $G := \{\1,\sigma\}$, which 
leads to the Klein-$4$-group 
\[ G_\tau := G \rtimes \{\1,\tau\} \cong \Z/2\Z \times \Z/2\Z.\] 
We consider reflection positivity with respect to the subset 
$G_+ := \{\1\}$. 

Any unitary representation $(U,\cE)$ of $G_\tau$ decomposes into 
four eigenspaces 
\[\cE = \cE^{1,1} \oplus \cE^{-1,1} \oplus \cE^{1,-1} \oplus \cE^{-1,-1}, \qquad 
 \cE^{\eps_1, \eps_2} = \{ v \in\cE \: U_\sigma v = \eps_1 v, U_\tau v = \eps_2 v\},\] 
and for $\theta := U_\tau$, the subspace 
$\cE^\theta = \cE^{1,1} \oplus \cE^{-1,1}$ is $U_\sigma$-invariant. 
For $v = (a,b,c,d)$, we then have 
\[ U_\tau v = (a,b,-c,-d) \quad \mbox{ and }\quad 
  U_\sigma v = (a,-b,c,-d).\] 
Assume that $\cE_+ = \C v$ for a single vector~$v$. Then reflection positivity 
corresponds to 
\[ \la v,\theta v \ra = |a|^2 + |b|^2 - |c|^2 - |d|^2 \geq 0.\] 
With respect to $U_\sigma$, we have 
\[ v = v_1 + v_{-1} = (a,0,c,0) + (0,b,0,d) \] 
and 
\[ \la U_\sigma v, \theta U_\sigma v \ra 
= \la v, \theta v \ra \geq 0 \quad \mbox{ and }\quad
 \la U_\sigma v, \theta  v \ra 
= |a|^2 - |b|^2 - |c|^2 + |d|^2.\] 
Therefore the subspace $\C v + \C U_\sigma v$ is $\theta$-positive if and  only if 
\[ \pm(|a|^2 - |b|^2 - |c|^2 + |d|^2\big) \leq |a|^2 + |b|^2 - |c|^2 - |d|^2,\] 
which is equivalent to 
\[ |d| \leq |b| \quad \mbox{ and }\quad |c| \leq |a|.\] 
Clearly, these two conditions are strictly stronger than the 
$\theta$-positivity of~$\C v$. 

For the corresponding positive definite function $f(g) = \la v, U_g v\ra$ we have
\begin{align*} 
f(\1) &= |a|^2 + |b|^2 + |c|^2 + |d|^2, \quad
&f(\tau) &= |a|^2 + |b|^2 - |c|^2 - |d|^2, \\ 
f(\sigma) &= |a|^2 - |b|^2 + |c|^2 - |d|^2, \qquad 
&f(\sigma\tau) &= |a|^2 - |b|^2 - |c|^2 + |d|^2. \\
\end{align*}
Decomposing $f = f_1 + f_{-1}$ with respect to the left action of $\sigma$, we obtain 
\[ f_1(\1) = f_1(\sigma) = |a|^2 + |c|^2, \qquad 
f_1(\tau) = f_1(\sigma\tau)  = |a|^2 -|c|^2 \] 
and 
\[ f_{-1}(\1) = -f_1(\sigma) = |b|^2 + |d|^2, \qquad 
f_{-1}(\tau) = -f_{-1}(\sigma\tau)  = |b|^2 -|d|^2. \] 
Both functions $f_{\pm 1}(g) = \la v_{\pm 1}, U_g v_{\pm 1}\ra$ 
are positive definite, but they are reflection positive if and only if
$|c| \leq |a|$ and $|d| \leq |b|$. 

Note that, even for $U_\sigma = \1$ and $U_\sigma = - \1$, there exist non-trivial 
reflection positive representations with $\la v, \theta v \ra > 0$.

\section{Reflection positive functions and KMS conditions} 

In this section we build the bridge from positive definite functions 
$\psi \: \R\to \Bil(V)$ 
satisfying the  KMS condition for $\beta > 0$ to reflection positive functions on 
the group $\T_{2\beta,\tau}\cong \OO_2(\R)$. We have already seen in Lemma~\ref{lem:2.2} that analytic continuation leads to a $2\beta$-periodic function 
$\phi \: \R\to \Bil(V)$ satisfying 
$\phi(t + \beta) =\oline{\phi(t)}$ for $t \in \R$ and 
$\phi(t) = \psi(it)$ for $0 \leq t \leq \beta$. In this section we show 
the existence 
of a positive definite function $f \: \R_\tau \to \Bil(V)$ with 
$f(t,\tau) = \phi(t)$ for $t \in \R$. By construction, $f$ is then reflection 
positive with respect to the interval $[0,\beta/2] = G_+ \subeq G = \R$ 
in the sense of Definition~\ref{def:1.2c}. 

Since we can build on Theorem~\ref{thm:kms}, our first goal is to express,  
for a standard real subspace $V \subeq \cH$, the $\Bil(V)$-valued function 
\begin{equation}
  \label{eq:phi1}
\phi \: [0,\beta]  \to \Bil(V), \qquad 
\phi(t)(v,w) := \psi(it)(v,w) = 
\la \Delta^{t/2\beta} v, \Delta^{t/2\beta} w \ra \quad \mbox{ for } \quad 
v,w \in V, 0 \leq t \leq \beta\end{equation}
from \eqref{eq:anacont1} in the proof of Theorem~\ref{thm:kms} 
as a $B(V_\C)$-valued function. 
To this end, we shall need the description 
of $V$ in terms of a skew-symmetric strict contraction $C$ on~$V$ 
(Lemma~\ref{lem:6.10}), and this 
leads to a quite explicit description of~$\phi$ 
that we then use to prove our main theorem.

\subsection{From form-valued to operator-valued functions} 

In the following it will be more convenient to work with 
operator-valued functions instead of form-valued ones. 
The translation is achieved by the following lemma. 
For its formulation,  we recall the polar decomposition of bounded 
skew-symmetric operators on real Hilbert spaces. 

\begin{rem} \mlabel{rem:6.11} (Polar decomposition of skew-symmetric operators) 
Let $D^\top = - D$ be an injective skew-symmetric operator on the 
real Hilbert space $V$ and let $D = I|D|$ be its polar decomposition. 
Then $\im(D)$ is dense because $D$ is injective, and therefore 
$I$ defines an isometry $V \to V$. From 
\[ I |D| =D =  - D^\top = - |D| I^{-1} = - I^{-1}(I|D| I^{-1}) \] 
it follows that $I^2 = -\1$, i.e., that $I$ is a complex structure and that 
$|D|$ commutes with $I$. 
\end{rem}

\begin{lem} \mlabel{lem:3.3} 
Let $V \subeq \cH$ be a standard real subspace with modular objects $(\Delta, J)$, 
let $\hat C := i \frac{\Delta - \1}{\Delta + \1}$, and 
let $C := \hat C\res_V \in B(V)$ 
be the skew-symmetric strict contraction from {\rm Lemma~\ref{lem:6.10}}. 
We assume that $\ker C = \{0\}$, so that 
the polar decomposition $C = I |C|$ defines a complex structure $I$ on~$V$. 
Consider the skew-symmetric operator 
\[ D :=  \log\Big(\frac{\1 -|C|}{\1 +|C|}\Big) I.\] 
Then the function $\phi(t)(v,w) = \la \Delta^{t/2} v, \Delta^{t/2} w \ra$ 
from \eqref{eq:phi1} has the form 
\begin{equation}
  \label{eq:phi-rel}
 \phi(t)(v,w) = \la v, \tilde\phi(t) w \ra_{V_\C} \quad \mbox{ for }\quad t \in [0,1], 
v,w \in V_\C,  
\end{equation}
where the function 
$\tilde\phi \: [0,1] \to B(V_\C)$ is given by 
\begin{align*}
 \tilde\phi(t) 
&= (\1 + i C)^{1-t} (\1- i C)^t 
= \frac{e^{-t|D|} + e^{-(1-t)|D|}} 
{\1 + e^{-|D|}}  + i I \frac{e^{-t|D|} - e^{-(1-t)|D|}}{\1 + e^{-|D|}}.
\end{align*}
\end{lem}

Note that 
$\tilde\phi(0) = \1 + i C \not= \1$ if $C \not=0$. 

\begin{prf} Since $C$ is a skew-symmetric contraction on $V$, the operators 
$\1 \pm i C$ on $V_\C$ are symmetric, so that we obtain a function  
\[\tilde\phi \:  [0,1] \to B(V_\C), \qquad 
\tilde \phi(t) := (\1 + i C)^{1-t} (\1 - i C)^t, \quad 0 \leq t \leq 1. \] 
Therefore both sides of \eqref{eq:phi-rel} are defined, and we have to show 
that 
\begin{equation}
  \label{eq:phirel-2}
\la v, \tilde\phi(t)w \ra_{V_\C} = \la \Delta^{t/2} v, \Delta^{t/2} w \ra 
\quad \mbox{ for} \quad v,w \in V_\C.
\end{equation}
For the skew-hermitian contraction $\hat C$ on $\cH$, we likewise obtain 
bounded operators 
\[ \hat \phi(t) := (\1 + i \hat C)^{1-t} (\1 - i \hat C)^t, \quad 0 \leq t \leq 1, \] 
and the continuity of the inclusion $V_\C \into \cH$ implies that 
\[ \hat\phi(t)\res_{V_\C}  = \tilde\phi(t) \:  V_\C \to V_\C.\] 

>From the relation 
$\Delta = \frac{\1 - i \hat C}{\1 + i \hat C}$ we further obtain the identity 
\[ \hat \phi(t) = (\1 + i \hat C) \Delta^t \] 
of selfadjoint operators on $\cH$. 
Let $V_\C'$ denote the domain of the (possibly) unbounded 
selfadjoint operator $\frac{\1 - i C}{\1 + i C}$ on $V_\C$. 
Then $V_\C'$ is a dense subspace which is contained in  the domain of 
$\big(\frac{\1 - i C}{\1 + i C}\big)^t$ for $0 \leq t \leq 1$.
For $w \in V_\C'$ we have 
\[ \tilde \phi(t) w = (\1 + i C) \Big(\frac{\1 - i C}{\1 + i C}\Big)^tw
\quad \mbox{ for } \quad 0 \leq t \leq 1.\] 
For $v \in V_\C$ and $\tilde w := \Big(\frac{\1 - i C}{\1 + i C}\Big)^t w$ we 
now obtain with \eqref{eq:h-rel} from Lemma~\ref{lem:6.10} the relation 
\begin{align*}
\la v , \tilde\phi(t) w \ra_{V_\C} 
&= \la v, (\1 + i C) \tilde w \ra_{V_\C} 
= \la v, \tilde w \ra_{\cH} 
= \la v, \Big(\frac{\1 - i C}{\1 + i C}\Big)^t w \ra_{\cH} \\
&= \la v, \Big(\frac{\1 - i \hat C}{\1 + i \hat C}\Big)^t w \ra_{\cH} 
= \la v, \Delta^t w \ra_{\cH} 
= \la \Delta^{t/2} v, \Delta^{t/2} w \ra_{\cH}.
\end{align*}
Since both sides of \eqref{eq:phirel-2} define continuous hermitian forms 
on $V_\C$ and the preceding calculation shows that equality holds on a dense subspace, 
we obtain \eqref{eq:phirel-2} for all $v,w \in V_\C$. 

Next we observe that the polar decomposition of $D$ is given by 
\[ D = - I|D| \quad \mbox{ and } \quad 
|D| = \log\Big(\frac{\1 +|C|}{\1 -|C|}\Big).\] 
The operator $|D|$ satisfies 
\begin{equation}
  \label{eq:deq}
e^{\mp |D|} = \frac{\1 \mp |C|}{\1 \pm |C|} \quad \mbox{ and } \quad 
1 + e^{\mp |D|} = \frac{2}{1 \pm |C|}. 
\end{equation}
Since $iI$ is an involution with the two eigenvalues $\pm 1$, comparing the action  
on both eigenspaces shows that, for $0 \leq t \leq 1$, we have 
\[ \Big(\frac{\1 - i C}{\1 + i C}\Big)^t
= \Big(\frac{\1 - i I|C|}{\1 + i I|C|}\Big)^t
= e^{-t|D| iI}.\] 
The assertion of the lemma now follows from 
\begin{align*}
\tilde \phi(t) 
&= (\1 + i C)\Big(\frac{\1 - i C}{\1 + i C}\Big)^t 
= (\1 + i I |C|) e^{-t|D| iI} 
= (\1 + i I |C|)\Big(e^{t|D|}\frac{\1 - i I}{2} +e^{-t|D|}\frac{\1 + i I}{2}\Big) \\ 
&= e^{-t|D|}(\1 + |C|)\frac{\1 + i I}{2} 
+  e^{t|D|}(\1 - |C|)\frac{\1 - i I}{2} 
= (\1 + |C|)\Big(e^{-t|D|} \frac{\1 + i I}{2} +  e^{-(1-t)|D|}\frac{\1 - i I}{2}\Big)\\
&= (\1 + e^{-|D|})^{-1} \Big(e^{-t|D|} (\1 + i I)+  e^{-(1-t)|D|}(\1 - i I)\Big)
= \frac{e^{-t|D|} + e^{-(1-t)|D|}}{\1 + e^{-|D|}}  
+ i I \frac{e^{-t|D|} - e^{-(1-t)|D|}}{\1 + e^{-|D|}}.
\qedhere\end{align*}
\end{prf}

\begin{rem} (a) 
Since $C$ is a strict contraction on $V$, $\1 + i C$ is injective on $V_\C$, so that 
\[ \tilde H 
:= \log\big(\frac{\1 + i C}{\1 - i C}\big) 
= \log\big(\frac{\1 + i I |C|}{\1 - i I |C|}\big) 
= iI |D| = - i D \] 
also defines a selfadjoint operator on the complex Hilbert space space $V_\C$. 

Next we observe that $\tilde H$ is a restriction of 
$L := \log\big(\frac{\1 + i \hat C}{\1 - i \hat C}\big) = - \log\Delta$, 
the infinitesimal generator of the one-parameter group
$U_t = \Delta^{-it}$. For the orthogonal one-parameter group 
$U^V_t := U_t\res_{V} \in \OO(V)$, it follows that its infinitesimal 
generator is a skew-adjoint extension of the skew-adjoint operator $D$ on~$V$, 
hence coincides with $D$. We therefore have 
\begin{equation}
  \label{eq:delts-d-rel}
e^{tD} = \Delta^{-it}\res_V \quad \mbox{ for } \quad t \in \R.
\end{equation}
This provides an alternative characterization of the operator $D$ 
in Lemma~\ref{lem:3.3}. 

(b) Let $(V, (\cdot,\cdot))$ be a real Hilbert space and $(U_t)_{t \in \R}$ 
be an orthogonal strongly continuous one-parameter group 
with skew-symmetric infinitesimal generator $D$, i.e., $U_t = e^{tD}$ for $t \in \R$. 
Let us assume that $\ker D = \{0\}$, resp., that the subspace 
$V^U$ of $U$-fixed points in $V$ is trivial. Then the polar decomposition 
$D = I |D|$ can be used to define a skew-symmetric contraction 
\[ C := (-I) \frac{\1 - e^{-|D|}}{\1 + e^{-|D|}} 
\quad \mbox{ with } \quad 
|C|= \frac{\1 - e^{-|D|}}{\1 + e^{-|D|}}.\] 
Then the hermitian form 
\[ h(v,w) :=  (v,w) + i (v,C w) \] 
defines a positive definite kernel on $V$ 
(Lemma~\ref{lem:1.3}). Let $\cH$ denote the corresponding reproducing kernel 
space and let $j \: V \to \cH$ be the natural map. 
By construction, $|C|$ has no fixed points, 
so that $\1 + C^2$ is injective, and therefore 
Lemma~\ref{lem:1.3}(iii) implies that the complex linear extension
$j_\C \: V_\C \to \cH$ is injective. As the real part of $h$ is 
the original scalar product on $V$, the inclusion 
$V \into\cH$ is isometric, so that $V \cong j(V)$ is a standard real subspace 
of $\cH$. Since $h$ is $U$-invariant, it defines a unitary 
one-parameter group $\hat U$ on $\cH$. 
Finally \eqref{eq:delts-d-rel} implies that 
$\hat U_t = \Delta^{-it}$ for $t \in \R$ and the modular operator 
$\Delta$ corresponding to $j(V)$. 
This shows that every orthogonal one-parameter group 
on a real Hilbert space is of the form~\eqref{eq:delts-d-rel} 
for a naturally defined embedding $V \into \cH$ as a standard real subspace. 
\end{rem}

Before we turn to the associated reflection positive functions, we need 
the following technical lemma on Fourier expansions of certain 
operator-valued functions. In \cite{CMV01} this is called the 
Matsubara formalism.\begin{footnote}{In view of \cite[Def.~18.49]{DG13}, 
we have 
\[  u_B^+(t) = G_{E,\beta}(t) \cdot \frac{2B(\1 - e^{-\beta B})}{\1 + e^{-\beta B}}, \] 
where $G_{E,\beta}$ is the euclidean thermal Green's function associated to the positive 
operator $\eps = B$.}  
\end{footnote}

\begin{lem} \mlabel{lem:four-exp} 
Let $B \geq 0$ be a selfadjoint operator on the complex 
Hilbert space $\cH$ and let $\beta > 0$. We consider the operator-valued functions 
$u^\pm_B\: \R \to B(\cH)$ 
satisfying 
\[ u^\pm_B(t + \beta) = \pm u^\pm_B(t) \quad \mbox{ and } \quad 
u^\pm_B(t) = \frac{e^{-tB} \pm e^{-(\beta -t)B}}{\1 + e^{-\beta B}}
\quad \mbox{ for } \quad 
0 \leq t \leq \beta.\] 
Then $u^\pm_B$ are weakly continuous symmetric $2\beta$-periodic 
with the Fourier expansions 
\[ u^+_B(t) = \sum_{n \in \Z} c_{2n}^B e^{ 2n \pi i t/\beta} \quad \mbox{ 
and } \quad 
u^-_B(t) = \sum_{n \in \Z} c_{2n+1}^B e^{ (2n+1)\pi i t/\beta} \] 
with 
\[ c_{n}^B = c_{-n}^B =  \frac{(\1 - (-1)^n e^{-\beta B})}{\1 + e^{-\beta B}}
\cdot \frac{2\beta B}{(\beta B)^2 + (n\pi)^2} \quad \mbox{ for } \quad n \in \Z
\] 
\end{lem}

\begin{prf} (a) Every $2\beta$-periodic continuous function 
$\xi \: \R \to \C$ has a Fourier expansion 
\[ \xi(t) = \sum_{n \in \Z} c_n e^{\pi i n t/\beta} 
\quad \mbox{ with } \quad 
c_n = \frac{1}{2\beta} \int_0^{2\beta} \xi(t) e^{-\pi i n t/\beta}\, dt.\] 
For the $\beta$-periodic 
function with $u^+(t) = u^+_\lambda(t) :=  \frac{e^{-t\lambda} + e^{-(\beta - t)\lambda}}{
1 + e^{-\beta\lambda}}$ for 
$0 \leq t \leq \beta$ we have $u^+(t+\beta) = u^+(t)$, so that only even terms appear: 
\[ u^+(t) = \sum_{n \in \Z} c_{2n} e^{\pi i 2n t/\beta}, \quad 
c_{2n}= \frac{1 - e^{-\beta \lambda}}{1 +  e^{-\beta \lambda}}
\frac{2\beta \lambda}{(\beta\lambda)^2 + (2\pi n)^2}.\] 
To obtain this formula, we first calculate 
\begin{align*}
a_{\lambda,n}
&:= \frac{1}{\beta} \int_0^{\beta} e^{-t\lambda} e^{-\pi i n t/\beta}\, dt
= \int_0^1 e^{-(\beta\lambda + \pi i n) t}\, dt
= \frac{1 - e^{-(\beta\lambda + \pi i n)}}{\beta \lambda + \pi i 2n}
= \frac{1 - (-1)^ne^{-\beta\lambda}}{\beta \lambda + \pi i 2n}.  
\end{align*}
Therefore 
\begin{align*}
(1 + e^{-\lambda\beta}) c_{2n} &= a_{\lambda,2n} + e^{-\beta\lambda} a_{-\lambda,2n}
=  \frac{1 - e^{-\beta\lambda}}{\beta \lambda + 2n\pi i} 
+ e^{-\beta\lambda}  \frac{1 - e^{\beta\lambda}}{-\beta \lambda + 2n\pi i} \\
&=  \frac{1 - e^{-\beta\lambda}}{\beta \lambda + 2n\pi i} 
+  \frac{1 - e^{-\beta\lambda}}{\beta \lambda - 2n\pi i}
=  \frac{(1 - e^{-\beta\lambda})2\beta \lambda}{(\beta \lambda)^2 + (2n\pi)^2}
\end{align*}

For the $2\beta$-periodic 
function with $u^-(t) = u^-_\lambda(t) :=  
\frac{e^{-t\lambda} - e^{-(\beta - t)\lambda}}{1 + e^{-\beta \lambda}}$ for 
$0 \leq t \leq \beta$ and $u^-(t+\beta) = -u^-(t)$ only odd terms appear: 
\[ u^-(t) = \sum_{n \in \Z} c_{2n+1} e^{\pi i (2n+1) t/\beta}, \quad 
c_{2n+1}=  \frac{2\beta\lambda}{(\beta \lambda)^2 + ((2n+1)\pi)^2}.\] 
This follows from 
\begin{align*}
c_{2n+1} &= \frac{a_{\lambda,2n+1} - e^{-\beta\lambda} a_{-\lambda,2n+1}}
{1 + e^{-\beta\lambda}}
=  \frac{1}{\beta \lambda + (2n+1)\pi i} 
- \frac{e^{-\beta\lambda}(1 + e^{\beta\lambda})}{1 + e^{-\beta\lambda}}
\frac{1}{-\beta \lambda + (2n+1)\pi i} \\
&=  \frac{1}{\beta \lambda + (2n+1)\pi i} 
+  \frac{1}{\beta \lambda - (2n+1)\pi i}
=  \frac{2\beta\lambda}{(\beta \lambda)^2 + ((2n+1)\pi)^2}
\end{align*}
Note that 
\[ c_{n} = c_{-n} =  \frac{1 - (-1)^n e^{-\beta\lambda}}{1 + e^{-\beta\lambda}}
\frac{2\beta \lambda}
{(\beta \lambda)^2 + (n\pi)^2} \quad \mbox{ for } \quad n \in \Z.\] 

(b) If $P$ denotes the spectral measure of $B$, we have for $v \in \cH$ the relation 
\[ \la v,Bv \ra = \int_0^\infty x \, dP^{v,v}(x) \quad \mbox{ with } \quad 
P^{v,v} = \la v, P(\cdot)v\ra.\] 
This leads for $0 \leq t \leq 2 \beta$ to 
\[ \la v,u^\pm_B(t)v \ra = \int_0^\infty u^\pm_\lambda(t)\, dP^{v,v}(\lambda). \]
For the operator-valued Fourier coefficients, we thus obtain 
\begin{align*}
\la v,c_n^Bv \ra 
&= \int_\R c_n(\lambda)\, dP^{v,v}(\lambda)
= \int_\R \frac{1 - (-1)^n e^{-\beta\lambda}}{1 + e^{-\beta\lambda}}
\frac{2\beta \lambda}
{(\beta \lambda)^2 + (n\pi)^2} \, dP^{v,v}(\lambda)\\
&= \Big \la v, \frac{(\1 - (-1)^n e^{-\beta B})}{\1 + e^{-\beta B}}
\frac{2\beta B}{(\beta B)^2 + (n\pi)^2}v\Big\ra.
\end{align*}
This proves the assertion. 
\end{prf}

\subsection{Existence of reflection positive extensions} 

We now come to one of our main result on reflection positive extensions. 
It shows that, for every positive definite function 
$\psi \: \R \to \Bil(V)$ satisfying the $\beta$-KMS condition, 
there exists a reflection positive function 
$f \: G_\tau \to B(V_\C)$ satisfying 
$\psi(it)(v,w) = \la v, f(it,\tau)w \ra$ for $v,w \in V, 0 \leq t \leq \beta$. 
Then the corresponding GNS representation 
$(U^f, \cH_f)$ of the group $(\T_{2\beta})_\tau \cong \OO_2(\R)$ 
is a ``euclidean realization'' 
of the unitary one-parameter group 
$(\Delta^{-it/\beta})_{t \in \R}$ corresponding to~$\psi$ in the sense 
that it is obtained by Osterwalder--Schrader quantization 
from $U^f$ (cf.\ \cite{NO14a}). The following theorem generalizes 
the results of \cite{NO15} dealing with the scalar-valued case. 

\begin{thm} \mlabel{thm:2} {\rm(Reflection positive extensions)}  
Let $V \subeq \cH$ be a standard real subspace and let 
$C = I|C|$ be the corresponding skew-symmetric strict contraction on $V$. 
We assume that $\ker C = \{0\}$, so that $I$ defines a complex structure on~$V$.
We define a weakly continuous function 
$\tilde \phi \:\R \to B(V_\C)$ by
\[ \tilde\phi(t) = (\1 + i C)^{1-t/\beta} (\1 - i C)^{t/\beta} \quad \mbox{ for } \quad 
0 \leq t \leq \beta \quad \mbox{ and } \quad 
\tilde\phi(t + \beta) = \oline{\tilde\phi(t)}\quad \mbox{ for } \quad t \in \R.\] 
Write 
\[ \tilde\phi(t) = u^+(t) + i I u^-(t)\quad \mbox{ with } \quad 
u^\pm(t) \in B(V), \quad u^\pm(t+ \beta) = \pm u^\pm(t).\] 
Then 
\[ f \: \R_\tau \to B(V_\C), \qquad 
f(t,\tau^\eps) := u^+(t) + (iI)^\eps u^-(t), \qquad t \in \R, \eps \in \{0,1\}, \] 
is a weak-operator continuous positive definite function satisfying 
$f(t,\tau) = \tilde\phi(t)$. 
It is reflection positive with respect to the subset $[0,\beta/2] \subeq \R$ 
in the sense that the kernel 
\[ f\big((t,\tau)(-s,\1)\big) = f(t + s,\tau), \qquad 0 \leq s,t \leq \beta/2 \] 
is positive definite. 
\end{thm}

\begin{prf} We may w.l.o.g.\ assume that $\beta = 1$. 
Recall the operator $D$ from Lemma~\ref{lem:3.3}. 
With this lemma, we write 
\begin{align*}
 \tilde\phi(t) 
&= (\1 + e^{-|D|})^{-1}\Big(e^{-t|D|} + e^{-(1-t)|D|} + i I (e^{-t|D|} - e^{-(1-t)|D|})\Big) 
\quad \mbox{ for } \quad 0 \leq t \leq 1.
\end{align*}
Using Lemma~\ref{lem:four-exp} with $\beta = 1$ and $B = |D|$, we get 
\[  \tilde\phi(t) 
= u^+_{|D|}(t) + i I u^-_{|D|}(t) \quad \mbox{ for } \quad 
t \in \R.\] 

(a) We define $f_1 \: \R_\tau \to B(V_\C)$ by 
$f_1(t, \tau^\eps) := u^+_{|D|}(t)$ for $t \in \R, \eps \in \{0,1\}$. 
To see that $f_1$ is positive definite, it suffices to verify this 
for its restriction to $\R$ (Lemma~\ref{lem:biinvar}), which follows 
from the positivity of the Fourier coefficients in the expansion 
\[ u^+_{|D|}(t) = \sum_{n \in \Z} c_{2n}^{|D|} e^{ 2n \pi i t} \quad \mbox{ 
with } \quad 
c_{2n}^{|D|} =  \frac{\1 - e^{-|D|}}{\1 + e^{-|D|}}
\frac{2 |D|}
{|D|^2 + (2n\pi)^2\1} \geq 0\] 
(Lemma~\ref{lem:four-exp}). Note that $f_1$ is $1$-periodic. 

(b) Likewise, the function $f_2 \: \R_\tau \to B(V_\C)$ defined by 
$f_2(t,\tau^\eps) := u^-_{|D|}(t)$ for $t \in \R, \eps \in \{0,1\}$ 
is positive definite because the 
Fourier coefficients 
\[ c_{2n+1}^{|D|} =  \frac{2 |D|}
{|D|^2 + ((2n+1)\pi)^2\1} \geq 0 \quad \mbox{ for } \quad n \in \Z\] 
are positive. Note that $f_2(t+1, \tau^\eps) = - f_2(t,\tau^\eps)$ for 
$t \in \R, \eps \in \{0,1\}$. 

(c) We now consider the function 
\[ \tilde f_2(g) := h(g) f_2(g) \quad \mbox{ with }\quad 
h(t,\tau^\eps) = (i I)^\eps \quad \mbox{ for } \quad t \in \R, 
\eps \in \{0,1\}. \] 
Since $h(g)$ commutes with $f_2(g')$ for $g,g'\in \R_\tau$, 
the function $\tilde f_2$ is positive definite if 
$h$ is positive definite (Lemma~\ref{lem:3.4}). 
As $h$ is constant on the two $\R$-cosets and its restriction 
to the $2$-element subgroup $\{\1,\tau\}$ is a unitary representation, 
$h$ is positive definite. 
We conclude that the $B(V_\C)$-valued function 
$f:= f_1 + \tilde f_2$ on $\R_\tau$ is positive definite. 
\end{prf}

\begin{cor} Let $V$ be a real vector space and let 
$\psi \: \R \to \Bil(V)$ be a continuous positive definite function 
satisfying the $\beta$-KMS condition. Then there exists a 
pointwise continuous function $f \: \R_\tau \to \Bil(V)$ 
which is reflection positive with respect to the subset 
$[0,\beta/2] \subeq \R$ and which satisfies 
\[ f(t,\tau) = \psi(it) \quad \mbox{ for } \quad 0 \leq t \leq \beta \qquad 
\mbox{ and } \quad 
f(t+\beta,\tau) = \oline{f(t,\tau)} \quad \mbox{ for } \quad t\in \R.\] 
\end{cor}

\begin{rem} The function $\tilde f_2$ in the proof 
of Theorem~\ref{thm:2} is  not reflection positive because 
$\tilde f_2(\beta,\tau)$ is a negative operator. This 
also shows that the natural decomposition 
$f = f_1 + \tilde f_2$ into even and odd part is not compatible 
with reflection positivity. 
\end{rem}

\subsection{Integral representation of reflection positive functions} 

We now describe an integral representation 
of the reflection positive function $f \: \R_\tau \to \Bil(V)$ which 
corresponds to the decomposition of the corresponding 
unitary representation of~$\R_\tau$. 
With 
\[ \tilde\phi(t) = (\1 + i C)^{1- t/\beta}(\1 - i C)^{t/\beta} \quad \mbox{ for } \quad 
0 \leq t \leq \beta, \] 
where $C \in B(V)$ is a skew-symmetric strict contraction, we 
first decompose $V$ into $V_0 := \ker C$ and $V_1 := V_0^\bot = \oline{CV}$. 
Then the polar decomposition $C = I |C|$ yields a complex structure 
$I$ on~$V_1$. Accordingly, we write 
$\tilde\phi = \tilde\phi_0 + \tilde\phi_1$, where 
$\tilde\phi_0 = \1$ is constant. This component leads to the 
constant function $f_0(t,\tau) =\1$. 
We now assume that $V = V_1$, i.e., that $C$ is injective. Then 
$I$ is a complex structure on $V$. 

\begin{prop} If $\ker C = \{0\}$ and $P$ denotes the spectral measure of 
the symmetric operator 
$|D| = \frac{1}{\beta}\log\frac{\1 + |C|}{\1 - |C|}$ on $V$, 
then we have the integral representation 
\begin{equation}
  \label{eq:intrep-f}
f(t,\tau^\eps) = \int_{(0,\infty)} 
u^+_\lambda(t) + u^-_\lambda(t) (iI)^\eps\, dP(\lambda), 
\end{equation}
where $u^\pm_\lambda \: \R \to \R$ are defined by 
$u^\pm_\lambda(t + \beta) = \pm u^\pm_\lambda(t)$ 
and 
\[ u^\pm_\lambda(t) :=  \frac{e^{-t\lambda} \pm e^{-(\beta - t)\lambda}}{
1 + e^{-\beta\lambda}}  \quad \mbox{ for }\quad 
0 \leq t \leq \beta.\] 
\end{prop}

\begin{prf} First we observe that 
$|D|$ is a positive 
symmetric operator with trivial kernel which commutes with~$I$. 
We therefore have 
$|D| = \int_{(0,\infty)} \lambda\, dP(\lambda)$. 
With the notation from Lemma~\ref{lem:four-exp}, we then have 
\[ f(t,\tau^\eps) = u^+_{|D|}(t) + u^-_{|D|}(t) (iI)^\eps \quad \mbox{ for } 
\quad t \in \R, \eps \in \{0,1\}.\] 
>From the integral representations 
$u^\pm_{|D|}(t) = \int_{(0,\infty)} u^\pm_\lambda(t)\, dP(\lambda),$ 
we now obtain \eqref{eq:intrep-f}.   
\end{prf}

\begin{rem} (a) For $0 \leq t \leq \beta$, we have in particular 
\[ f(t,\tau^\eps) = \int_{(0,\infty)} 
\frac{e^{-t\lambda} + e^{-(\beta - t)\lambda}}{1 + e^{-\beta\lambda}} \1 
+ \frac{e^{-t\lambda} -  e^{-(\beta - t)\lambda}}{1 + e^{-\beta\lambda}} (iI)^\eps
\, dP(\lambda).\] 

(b) The most basic type is obtained for 
$D = \lambda \1$, $ \lambda > 0$, which leads to 
\[ f(t,\tau^\eps) = \frac{(e^{-t\lambda} + e^{-(\beta - t)\lambda}\big) \1 + 
\big(e^{-t\lambda} -  e^{-(\beta - t)\lambda}\big) (iI)^\eps}{1 + e^{-\beta\lambda}} 
= u^+_\lambda(t)\1 + u^-_\lambda(t) (iI)^\eps \quad \mbox{ for }\quad 0 \leq t \leq \beta.\] 
The simplest non-trivial example arises for $V = \R^2$ with 
$I = \pmat{0& - 1\\ 1 & 0}$. 

(c) Every Borel spectral measure $P$ on $(0,\infty)$ which commutes with $I$ 
defines a positive operator $|D| = \int_0^\infty \lambda\, dP(\lambda)$ 
and we may put $D :=  -I |D|$. 
Then $\ker |D| = 0$, so that 
\[ |C| := \frac{e^{\beta |D|} - \1}{e^{\beta |D|} + \1} 
= \tanh\Big(\frac{\beta |D|}{2}\Big)\] 
is a symmetric contraction with trivial kernel commuting with $I$, 
and therefore $C := I|C|$ is a skew-symmetric contraction with polar 
decomposition $I|C|$ and $|D| = \frac{1}{\beta} \log\big(\frac{\1 + |C|}{\1 - |C|}\big)$.    
\end{rem}

\subsection{Characterizing reflection positive extensions} 
\mlabel{subsec:4.4}

In Theorem~\ref{thm:2} we obtained positive definite 
extensions to all or $\R_\tau$ for certain functions 
on the coset $\R \rtimes \{\tau\}$. 
In this section we describe an intrinsic characterization 
of those weakly continuous reflection positive functions 
$f \: \R_\tau \to B(V_\C)$ arising from this construction.
First we observe that we can recover $\psi$ from $f$: 

\begin{lem} If $f \: \R_\tau \to \Bil(V)$ is reflection positive and pointwise 
continuous, 
then there exists a unique $\beta$-KMS positive definite function 
$\psi \: \R \to \Bil(V)$ with 
\[ f(t,\tau) = \psi(it) \quad \mbox{ for } \quad 0 \leq t \leq \beta.\] 
\end{lem}

\begin{prf} First we observe that the function 
$\phi(t) := f(t,\tau)$ has values in $\Herm(V_\C)$ and satisfies 
\begin{equation}
  \label{eq:transrel}
\phi(t + \beta) = \oline{\phi(t)} \quad \mbox{ for } \quad t \in \R. 
\end{equation}
Reflection positivity implies that the kernel 
$\phi\big(\frac{t+s}{2}\big)$ for $0 \leq t,s \leq \beta$ 
is positive definite. By  \cite[Thm.~B.3]{NO15}, 
there exists a $\Bil^+(V)$-valued measure $\mu$ such that 
\begin{equation}
  \label{eq:lapl-form}
\phi(t) = \int_\R e^{-\lambda t}\, d\mu(\lambda) 
\quad \mbox{ for }\quad 
0 < t < \beta.
\end{equation}
The continuity of $\phi$ on $[0,\beta]$ actually implies that the 
integral representation also holds on the closed interval $[0,\beta]$ 
by the Monotone 
Convergence Theorem. In particular, the measure $\mu$ is finite. 
Therefore its Fourier transform 
$\psi(t) := \int_\R e^{it\lambda}\, d\mu(\lambda)$ is a pointwise continuous 
$\Bil(V)$-valued positive definite function on $\R$. 
Further, \eqref{eq:transrel} implies 
\begin{equation}
  \label{eq:measrel}
e^{\beta\lambda}\, d\mu(-\lambda) = d\oline{\mu}(\lambda)
\end{equation}
and Theorem~\ref{thm:2} shows that $\phi(t) = \psi(it)$ holds for the 
$\beta$-KMS function $\psi \: \R \to \Bil(V)$.
\end{prf}

Before we describe a realization of the GNS 
representation  $(U^f, \cH_f)$ in spaces of sections of a 
vector bundle, let us recall the general background for this. 

\begin{rem}
 For a $B(V)$-valued positive definite function 
$f \: G \to B(V)$, the reproducing kernel Hilbert space 
with kernel $K(g,h) = \phi(gh^{-1}) = K_g K_h^*$ is generated by the functions 
\[ K_{h,w} := K_h^* w \quad \mbox{ with } \quad 
K_{h,w}(g) = K_g K_h^* w = K(g,h) w = \phi(gh^{-1})w.\] 
The group  $G$ acts on this space by right translations 
\[  (U_g s)(h) := s(hg).\] 

If $P \subeq G$ is a subgroup and $(\rho,V)$ is a unitary representation 
for which 
\[ f(hg) = \rho(h) f(g) \quad \mbox{ for all } \quad g\in G, h \in P,\] 
then 
\[ \cH_f \subeq \cF(G,V)_\rho := \{ s \: G \to V \:  (\forall g \in G)(\forall h \in P)\, 
s(hg) = \rho(h) s(g)\}.\] 
Therefore $\cH_f$ can be identified with a space of sections of the 
associated bundle 
\[ \V := (V \times_P G) = (V \times G)/P,\] where
$P$ acts on the trivial vector bundle $V \times  G$ over $G$ by 
$h.(v,g) = (\rho(h)v, hg)$. 
\end{rem} 

To derive a suitable characterization of the functions 
$f$ arising in Theorem~\ref{thm:2}, we identify 
$2\beta$-periodic function $s$ on $\R$ with pairs of function 
$(s_0, s_1)$ via $s = s_0 + s_1$, where $s_0$ is $\beta$-periodic and 
$s_1(\beta + t)= - s_1(t)$. 
Accordingly, any $2\beta$-periodic function 
$s \: \R \to V_\C$ defines a function
\[  \tilde s \: \R \to V_\C^2, \quad \tilde s = (s_1, s_2) 
\quad \mbox{ with }\quad 
\tilde s(\beta + t) = \pmat{\1 & 0 \\ 0 & -\1} \tilde s(t).\] 
In this sense $\tilde s$ is a section of the vector bundle 
over $\T_\beta$ with fiber $V_\C^2$ defined by the representation of 
$\beta\Z$, specified by 
\[ \rho(\beta) = \pmat{\1 & 0 \\ 0 & -\1}.\] 

Splitting the $B(V)$-valued positive definite function 
\[f \: \R_\tau \to B(V), \qquad 
f(t,\tau^\eps) =u^+_{|D|}(t) + u^-_{|D|}(t) (iI)^\eps 
\quad \mbox{ for }\quad t \in \R, \eps \in \{0,1\}\] 
into even and odd part with respect to the $\beta$-translation, we obtain: 

\begin{lem} \mlabel{lem:indrep} For the 
subgroup $P := (\Z\beta)_\tau \cong \Z\beta \rtimes \{\1,\tau\}$ of 
$G := \R_\tau$, we consider the unitary representation $\rho \: P \to \U(V_\C^2)$ defined by 
\[ \rho(\beta,\1) := \pmat{ \1 & 0 \\ 0 & -\1} \quad \mbox{ and } \quad 
\rho(0,\tau) := \pmat{ \1 & 0 \\ 0 & i I},\] 
where $I$ is a complex structure on the real Hilbert space 
$V$ commuting with the positive operator $|D|$. 
Then 
\[ f^\sharp \: \R_\tau \to B(V^2) \cong M_2(B(V)), \qquad 
f^\sharp(t, \tau^\eps) := \pmat{ u^+_{|D|}(t) & 0 \\ 0 & u^-_{|D|}(t) (iI)^\eps} \] 
is a positive definite function satisfying 
\begin{equation}
  \label{eq:covar-f}
f^\sharp(hg) = \rho(h) f^\sharp(g) \quad \mbox{ for } \quad 
h \in P, g \in G.
\end{equation}
The corresponding GNS representation $(U^{f^\sharp}, \cH_{f^\sharp})$ 
is equivalent  to the 
GNS representation $(U^f, \cH_f)$. 
\end{lem}

\begin{prf} The first assertion follows from 
\[ f^\sharp((0,\tau)(t,\tau^\eps)) = f^\sharp(-t, \tau^{\eps +1}) 
= \pmat{u^+_{|D|}(-t) & 0 \\ 0 & u^-_{|D|}(-t) (iI)^{\eps + 1}} 
= \pmat{u^+_{|D|}(t) & 0 \\ 0 & u^-_{|D|}(t) (iI)^{\eps + 1}} \] 
and 
\[ f^\sharp(\beta + t, \tau^\eps) 
= \pmat{u^+_{|D|}(t) & 0 \\ 0 & - u^-_{|D|}(t) (iI)^{\eps}}.\] 
As the GNS representation $(U^f, \cH_f)$ decomposes under the involution 
$U^f_\beta$ into $\pm 1$-eigenspaces, this representation is equivalent to 
the GNS representation $(U^{f^\sharp}, \cH_{f^\sharp})$ corresponding to~$f^\sharp$. 
\end{prf}

\begin{rem} \mlabel{rem:4.13} (a) The preceding lemma implies that, 
if the complex structure $I$ on $V$ is fixed, then 
the relation \eqref{eq:covar-f} determines $f^\sharp$ completely in terms 
of the function 
\[ [0,\beta]\to M_2(B(V)), \quad t \mapsto f^\sharp(t,\tau) 
= \pmat{ \Re\phi(t) & 0 \\ 
0 & i \Im \phi(t)},\] 
so that $\phi$ determines~$f$ in a natural way.  

(b) This lemma also shows that
we may identify the Hilbert space $\cH_f \cong \cH_{f^\sharp}$ 
 as a space of section of a Hilbert bundle $V^2 \times_\rho G$ 
over the circle $\T_\beta \cong \R_\tau/H$ with fiber $V^2$. 

(c) Every function $s \: \R_\tau \to V^2$ satisfying 
$s(hg) = \rho(h) s(g)$ for $h \in (\beta \Z)_\tau$ is determined by 
its restriction $\tilde s$ to the subgroup $\R$, which satisfies 
\[ \tilde s(\beta + t) = \rho(\beta,\1) \tilde s(t) \quad \mbox{ for }\quad t \in \R.\] 
The action of $\tau$ is in this picture given by 
\begin{equation}
  \label{eq:tau-act}
(\tau.\tilde s)(t) := s(t,\tau) = s((0,\tau)(-t,\1)) 
= \rho(\tau) \tilde s(-t).
\end{equation}
\end{rem}

\begin{rem} (a) In view of \eqref{eq:measrel}, there exists a 
$\Bil^+(V)$-valued measure $\nu$ on $[0,\infty)$ 
for which we can write 
\begin{equation} \label{eq:sum-mu}
d\mu(\lambda) = d\nu(\lambda) + e^{\beta\lambda} d\oline{\nu}(-\lambda). 
\end{equation}
This leads for $0 \leq t \leq \beta$ and $\nu = \nu_1 + i \nu_2$ to 
\begin{equation}
  \label{eq:phi-intrep}
\phi(t) 
= \int_0^\infty e^{-t\lambda} + e^{-(\beta - t)\lambda}\, d\nu_1(\lambda) 
 + i \int_0^\infty e^{-t\lambda} - e^{-(\beta - t)\lambda}\, d\nu_2(\lambda). 
\end{equation}

In particular, the most elementary non-trivial examples correspond to the 
Dirac measures of the form $\nu 
= \delta_\lambda \cdot (\gamma+ i \omega)$, where $\delta_\lambda$ is the Dirac measure 
in $\lambda > 0$: 
\[ \phi(t) = (e^{-t\lambda} + e^{-(\beta - t)\lambda})\gamma 
+ i (e^{-t\lambda} - e^{-(\beta - t)\lambda})\omega
= e^{-t\lambda} h  + e^{-(\beta - t)\lambda} \oline h,\quad \mbox{ where } \quad 
h := \gamma + i\omega \in \Bil^+(V).\] 

Writing $\omega(v,w) = \gamma(v,Cw)$ (Corollary~\ref{cor:1.3}) 
and replacing $V$ by the real Hilbert space 
defined by the positive semidefinite form $\gamma$ on $V$, we obtain 
the $B(V_\C)$-valued function 
\[ \tilde\phi(t) 
= (e^{-t\lambda} + e^{-(\beta - t)\lambda})
+ i C(e^{-t\lambda} - e^{-(\beta - t)\lambda})
= e^{-t\lambda}(\1 + i C) + e^{-(\beta - t)\lambda}(\1 - i C) \quad 
\mbox{ for } \quad 0 \leq t \leq \beta,\] 
which leads to 
\[f(t,\tau^\eps) 
=(1 + e^{-\beta\lambda}) (u^+_\lambda(t) \1 + u^-_\lambda(t) |C| (iI)^\eps)
\quad \mbox{ for }\quad t \in \R, \eps \in \{0,1\}.\] 

(b) This can also be formulated in terms of forms. With 
$\gamma(v,w) = \la v,w \ra_V$ and 
\[ h(v,w) = \gamma(v,w) + i \omega(v,w) 
= \la v,(\1 + i C)w\ra_{V_\C} = \la v,(\1 + i I |C|)w\ra_{V_\C},\] 
we get 
\[ f(t,\tau^\eps)(v,w) =\la v, (u^+_\lambda(t) \1 + u^-_\lambda(t) |C| (iI)^\eps)w\ra.\]
\end{rem}

\subsection{Realization by resolvents of the Laplacian} 

We have seen in the preceding subsection 
how to obtain a realization of the Hilbert space 
$\cH_{f}$  
as a space $\cH_{f^\sharp}$ 
of sections of a Hilbert bundle $\bV$ with fiber $V_\C^2$ over 
the circle $\T_\beta = \R/\beta \Z$. In this section we provide an 
analytic description of the scalar product on this space 
if $|D| = \lambda \1$ for some $\lambda > 0$. 
We shall see that it has a natural description 
in terms of the resolvent $(\lambda^2 - \Delta)^{-1}$ of the 
Laplacian of $\T_\beta$ acting on section of the bundle~$\bV$. 

On the circle group $\T_{2\beta}$, we consider the normalized 
Haar measure given by 
\[ \int_{\T_{2\beta}} h(t)\, d\mu_{\T_{2\beta}} 
= \frac{1}{{2\beta}} \int_0^{2\beta} h(t)\, dt,\] 
where we identify functions $h$ on $\T_{2\beta}$ with ${2\beta}$-periodic functions  
on~$\R$. 
 

As in Lemma~\ref{lem:indrep}, we write 
\[ f^\sharp(t,\tau^\eps) = \pmat{ u^+_\lambda(t) \1  & 0 \\ 0 & 
u^-_\lambda(t) (iI)^\eps} \in B(V_\C^2) \cong M_2(B(V_\C)),\] 
For $\chi_n(t) = e^{\pi i n t/\beta}$ we then have 
\[ u_\lambda^+ = \sum_{n \in \Z} c_{2n}^\lambda \chi_{2n} \quad  \mbox{ and } 
\quad u_\lambda^- = \sum_{n \in \Z} c_{2n+1}^\lambda \chi_{2n+1}, \] 
where 
\[ c_{n}^\lambda = c_{-n}^\lambda 
=  \frac{1 - (-1)^n e^{-\beta \lambda}}{1 + e^{-\beta\lambda}}
\cdot \frac{2\beta\lambda}{(\beta \lambda)^2 + (n\pi)^2} 
=  \frac{1 - (-1)^n e^{-\beta \lambda}}{1 + e^{-\beta\lambda}}
\cdot \frac{2\lambda}{\beta} \cdot \frac{1} 
{\lambda^2 + (n\pi/\beta)^2} \quad \mbox{ for } \quad n \in \Z \] 
(the rightmost factors are called bosonic 
Matsubara coefficients if $n$ is even and fermionic if $n$ is odd 
\cite[\S 18]{DG13}). 
With 
\begin{equation}
  \label{eq:cn-form2}
c^\lambda_+ := \frac{1 - e^{-\beta \lambda}}{1 + e^{-\beta\lambda}} 
\frac{2\lambda}{\beta} 
= \tanh\big(\frac{\beta\lambda}{2}\big) \frac{2\lambda}{\beta} 
\quad \mbox{ and }\quad 
c^\lambda_- := \frac{2\lambda}{\beta}, 
\end{equation}
we thus obtain 
\begin{equation}
  \label{eq:cn-form1}
c_{2n}^\lambda = \frac{c^\lambda_+}{\lambda^2 + (2n\pi/\beta)^2}, \qquad 
c_{2n+1}^\lambda = \frac{c^\lambda_-}{\lambda^2 + ((2n+1)\pi/\beta)^2}. 
\end{equation}

The following proposition shows that the 
positive  operator $(\lambda^2 - \Delta)^{-1}$  
on the Hilbert space of $L^2$-section of $\bV$ 
defines a unitary representation 
of $\R_\tau$ which is unitarily equivalent to the 
representation on $\cH_f$ (cf.~Lemma~\ref{lem:indrep}). 

\begin{prop} For $\lambda > 0$, let $\cH_\lambda$ be the Hilbert space obtained by completing the space 
\[ \Gamma_\rho := \{ s \in C^\infty(\R_\tau,V_\C^2) \: (\forall g \in 
\R_\tau, h \in (\Z \beta)_\tau)\ s(hg) =\rho(h)s(g)\} \] 
with respect to
\[ \la s_1, s_2 \ra := \frac{1}{2\beta}\int_0^{2\beta} 
\la s_1(t,\1), ((\lambda^2 - \Delta)^{-1} s_2)(t,\1) \ra \, dt. \] 
On $\cH_\lambda$ we have a natural unitary representation $U^\lambda$ 
of $\R_\tau$ by right translation which is unitarily equivalent to 
the GNS representation $(U^{f^\sharp}, \cH_{f^\sharp})$. Here the corresponding 
$j$-map is given by 
\begin{equation}
  \label{eq:jmap}
 j \: V \to \cH_\lambda, \quad 
j\pmat{v_1 \\ v_2} = 
\sqrt{c^\lambda_+} \sum_{n \in\Z} \chi_{2n} \pmat{v_1 \\ 0} 
+ \sqrt{c^\lambda_-} \sum_{n \in\Z} \chi_{2n+1} \pmat{0 \\ v_2}. 
\end{equation}
\end{prop}

\begin{prf} We identify $\Gamma_\rho$ with the space 
\[ \{ s \in C^\infty(\R,V_\C^2) \: (\forall t \in \R)\  
s(\beta + t) = \rho(\beta) s(t) \} \] 
(Remark~\ref{rem:4.13}).  
Then $s = \pmat{s_+ \\ s_-}$, where $s_+$ is $\beta$-periodic and 
$s_-$ is $\beta$-antiperiodic. 
Accordingly, we have an orthogonal decomposition 
$\cH_\lambda = \cH_\lambda^+ \oplus \cH_\lambda^-$, where 
$\cH_\lambda^\pm = \{ s \in \cH_\lambda \: (\forall t \in \R)\  
s(\beta + t) = \pm s(t)\}.$ 
Then $U^\lambda$ is given by 
\[ (U_t^\lambda s)(x) = s(t + x)\quad \mbox{ for } \quad t,x \in \R
\quad \mbox{ and }\quad (U_\tau^\lambda s)(x) = 
\pmat{ s_+(-x) \\ (iI) s_-(-x)}.\] 

>From the Fourier expansion $s = \sum_{n \in \Z} \chi_n s_{n}$ and the 
orthonormality of the $\chi_n$, we then derive 
\begin{equation}
  \label{eq:norm-form}
\la s_1,s_2\ra_{\cH_\lambda} 
= \sum_{n \in \Z} \frac{\la s_{1,n}, s_{2,n} \ra}{\lambda^2 + (n\pi/\beta)^2}.
\end{equation}
For the map $j \: V \to \cH_\lambda$ in \eqref{eq:jmap},  
the image is $U^\lambda_\R$-generating for $\cH_\lambda$ 
because the projection onto each Fourier component generates 
the first, resp., the second component of $V_\C^2$, according to parity. 
Therefore the unitary representation $(U^\lambda, \cH_\lambda)$ is 
equivalent to the GNS representation of the positive definite 
function $\tilde f \: \R_\tau \to B(V_\C^2)$, given by 
\[ \la \bv, \tilde f(g) \bw\ra = \la j(\bv), U^\tau_g j(\bw)\ra_{\cH_\lambda}.\] 
>From 
\[ U^\lambda_{(t,\tau^\eps)} j(\bv) = 
\sqrt{c^\lambda_+} \sum_{n \in\Z} \chi_{2n} \chi_{2n}(t) \pmat{v_1 \\ 0} 
+ \sqrt{c^\lambda_-} \sum_{n \in\Z} \chi_{2n+1} \chi_{2n+1}(t)\pmat{0 \\ (iI)^\eps v_2}, \] 
we derive with \eqref{eq:cn-form1} 
\begin{align*}
\la \bv, \tilde f(t,\tau^\eps) \bw \ra 
&= c^\lambda_+ \sum_{n \in\Z} 
\frac{\chi_{2n}(t)}{\lambda^2 + (2n\pi/\beta)^2} \la v_1, w_1 \ra 
+ c^\lambda_- \sum_{n \in\Z} 
\frac{\chi_{2n+1}(t)}{\lambda^2 + ((2n+1)\pi/\beta)^2} 
\la v_1, (iI)^\eps w_2 \ra \\
&=  \sum_{n \in\Z} 
\chi_{2n}(t) c^\lambda_{2n} \la v_1, w_1 \ra 
+ \sum_{n \in\Z} \chi_{2n+1}(t) c^\lambda_{2n+1} \la v_2, (iI)^\eps w_2 \ra \\
&=  \la v_1, u_\lambda^+(t) w_1 \ra 
+ \la v_2, u_\lambda^-(t)(iI)^\eps w_2 \ra 
= \la \bv, f^\sharp(t,\tau^\eps) \bw \ra. 
\end{align*}
This shows that $\tilde f = f^\sharp$, which completes the proof.
\end{prf}

\begin{rem}  From 
$u^+_\lambda = \sum_{n \in \Z} c_{2n}^\lambda \chi_{2n}$ 
it follows that 
\[ (\lambda^2- \Delta) u^+_\lambda 
=  \sum_{n \in \Z} c_{2n}^\lambda 
\Big(\lambda^2 + \frac{ (2\pi n)^2}{\beta^2}\Big) \chi_{2n} 
=  c^\lambda_+ \sum_{n \in \Z} \chi_{2n} 
= c^\lambda_+ \delta_0,\] 
where the latter relation means that 
\[ s_+(0) = \frac{1}{2\beta} \sum_{n \in \Z} \int_0^{2\beta} 
s_+(t) \chi_{2n}(t)\, dt \] 
for every smooth $\beta$-periodic functions $s_+$ on~$\R$. 
This relation can also be written as 
\[  (\lambda^2 - \Delta)^{-1} \delta_0 = \frac{1}{c^\lambda_+} u^+_\lambda.\] 

>From $u^-_\lambda  = \sum_{n \in \Z} c_{2n+1}^\lambda \chi_{2n+1}$ 
it follows that 
\[ (\lambda^2- \Delta) u^-_\lambda 
=  \sum_{n \in \Z} c_{2n+1}^\lambda 
\Big(\lambda^2 + \frac{ (2n+1)^2 \pi^2}{\beta^2}\Big) \chi_{2n+1}
=  c^\lambda_- \chi_1 \sum_{n \in \Z} \chi_{2n} \] 
As every smooth $\beta$-antiperiodic function $s_-$ is of the form 
$s_- = \chi_{-1} s_+$, where $s_+$ is $\beta$-periodic, we obtain, 
in the sense of distributions, 
\[  \la (\lambda^2- \Delta) u^-_\lambda, s_- \ra 
= c^\lambda_- s_+(0) = c^\lambda_- s_-(0) 
= \la c^\lambda_- \delta_0, s_- \ra,\] 
and therefore 
\[  (\lambda^2 - \Delta)^{-1} \delta_0 = \frac{1}{c^\lambda_-} 
u^-_\lambda\] 
on $\beta$-antiperiodic functions. 
Combining all this, we get 
\[ ((\lambda^2 - \Delta) f^\sharp)(t,\tau^\eps)
=  \pmat{(\lambda^2 - \Delta) u^+_\lambda \1 & 0 \\ 0 
&(\lambda^2 - \Delta) u^-_\lambda (iI)^\eps} 
= \delta_0 \pmat{c^\lambda_+ \1 & 0 \\ 0 & c^\lambda_- (iI)^\eps}\] 
as an operator-valued distribution on the space of smooth sections of~$\bV$ 
(cf.~also the discussion of thermal euclidean Green's 
functions in \cite[Def.~18.49]{DG13}).
\end{rem}

\section{The case $\beta = \infty$} 
\mlabel{sec:5} 

In the context of $C^*$-dynamical systems, it is well known that 
that the positive energy condition for the unitary one-parameter group 
implementing the automorphisms of a $C^*$-algebra $\cA$ in a representation 
can be viewed as a KMS condition for $\beta = \infty$ 
(cf.\ \cite{BR96}). For reflection positive representations of 
$G = \R$, this case corresponds to $G_+ = \R_+$, which has been treated in 
\cite{NO14a, NO14b} (cf.~also the discussion of euclidean Green's 
functions in \cite[Def.~18.48]{DG13}). 
The following theorem makes this analogy 
also transparent in the context of our Theorem~\ref{thm:kms}. 

If $\psi \: \R \to \Bil(V)$ is a positive definite function satisfying 
the KMS condition for $\beta > 0$, then its extension to 
$\oline{\cS_\beta}$ is pointwise bounded (Theorem~\ref{thm:kms}). 
This observation explains the assumptions in the following theorem. 

\begin{thm} \mlabel{thm:kms-infty} {\rm(KMS condition for $\beta = \infty$)} 
Let $V$ be a real vector space and 
let $\psi \:  \R \to \Bil(V)$ be a pointwise continuous positive definite function. 
Then the following are equivalent: 
\begin{itemize}
\item[\rm(i)] $\psi$ extends to a pointwise bounded function 
on the closed upper half plane which is pointwise holomorphic on $\C_+$. 
\item[\rm(ii)] There exists a $\Bil^+(V)$-valued regular Borel measure $\mu$ 
on $[0,\infty)$ satisfying 
\[ \psi(t) = \int_0^\infty e^{it\lambda}\, d\mu(\lambda).\] 
\item[\rm(iii)] The GNS representation $(U^\psi, \cH_\psi)$ has spectrum contained in 
$[0,\infty)$. 
\end{itemize}
If this is the case, then the function 
\[ f(t,\tau^\eps) := \psi(i|t|) \quad \mbox{ for }\quad t \in \R, \eps \in \{0,1\}, \] 
on $\R_\tau$ is reflection positive with respect to $\R_+ = [0,\infty)$. 
\end{thm}

\begin{prf} (i) $\Rarrow$ (ii): First we use \cite[Prop.~B.1]{NO15} to 
write $\phi$ as the Fourier transform of a $\Bil^+(V)$-valued regular 
Borel measure $\mu$ on $\R$: 
$\psi(t) = \int_\R e^{it\lambda}\, d\mu(\lambda).$ 
Evaluating in $v \in V_\C$, we obtain for the positive 
measure $\mu^{v,v} := \mu(\cdot)(v,v)$ the relation 
\[ \psi(t)(v,v) = \int_\R e^{it\lambda}\, d\mu^{v,v}(\lambda).\] 
This function extends to a bounded holomorphic function $\psi$ 
on $\C_+$. In particular, 
the Laplace transform $\cL(\mu^{v,v})(t) = \psi(it)(v,v)$ is bounded, which implies 
that $\supp(\mu^{v,v}) \subeq [0,\infty)$ (cf.\ \cite[Rem.~V.4.12]{Ne00}). 
This implies that $\mu$ is supported on $[0,\infty)$. 


(ii) $\Rarrow$ (iii): Write $U_t := U^\psi_t = e^{itH}$ with the selfadjoint 
generator $H$. We show that $H \geq 0$. 
Let $E$ be the spectral measure of $H$, so that 
$H = \int_\R \lambda\, dE(\lambda)$ and 
$U_t = \int_\R e^{it\lambda}\, dE(\lambda)$. 
It suffices to show that, for 
every $f \in L^1(\R)$ for which the Fourier transform 
$\hat f(\lambda) = \int_\R e^{i\lambda t}f(t)\, dt$ vanishes on $\R_+$, 
the operator 
\[ U_f 
= \int_\R f(t) e^{itH}\, dt 
= \int_\R \int_\R f(t) e^{it\lambda}\, dE(\lambda)\, dt 
= \int_\R \int_\R f(t) e^{it\lambda}\, dt \, dE(\lambda)
= \int_\R \hat f(\lambda)\, dE(\lambda) = \hat f(H) \] 
vanishes. 
For $v,w \in V$, we obtain with (ii) that 
\begin{align*}
 \la j(v), U_f j(w) \ra 
&= \int_\R f(t)\la j(v), U_t j(w) \ra\, dt  
= \int_\R f(t) \int_0^\infty e^{it\lambda}\, d\mu^{v,w}(\lambda)\, dt \\
&= \int_0^\infty \int_\R f(t) e^{it\lambda}\, dt\, d\mu^{v,w}(\lambda) 
= \int_0^\infty \hat f(\lambda)\, d\mu^{v,w}(\lambda) =0
\end{align*}
if $\hat f$ vanishes on $\R_+$. This proves that 
$j(V) \subeq \ker(U_f)$ and since $U_f$ is an intertwining operator 
and the subspace $j(V) \subeq \cH_\psi$ is generating, it follows 
that~$U_f =0$. This implies that $H \geq 0$. 

(iii) $\Rarrow$ (i): Write $U_t := U^\psi_t = e^{itH}$ and assume that $H \geq 0$. 
The spectral calculus for selfadjoint operators now implies that 
$\hat U_z := e^{izH}$, $\Im z \geq 0$ defines a 
strongly continuous representation on the upper half plane 
$\C_+$ which is holomorphic on the interior and whose range consists of 
contractions (\cite[Ch.~VI]{Ne00}). Then 
\[ \hat\psi(z)(v,w) 
= \la j(v), \hat U_z  j(w) \ra = 
 \la j(v), e^{izH} j(w) \ra, \qquad v, w\in V, \Im z \geq 0 \] 
provides the bounded analytic extension of $\psi$ to $\C_+$. 

Now we assume that (i)-(iii) are satisfied. 
Writing $\psi(t)(v,w) = \la j(v), U_t  j(w) \ra$ for a linear map \break 
$j \: V \to \cH$ and a unitary one-parameter group $U_t = e^{itH}$ on $\cH$, 
we have $H \geq 0$ by (iii) and 
\[ f(t, \tau^\eps) = \la j(v), e^{-|t|H} j(w) \ra,\] 
so that the positive definiteness of $f$ follows from the 
positive definiteness of the function $t \mapsto e^{-|t|H}$ on $\R$ 
(\cite[Prop.~4.1]{NO14a}). 
\end{prf}

\appendix

\section{Some background on positive definite kernels} 
\mlabel{sec:a}

In this appendix we collect precise statements of 
some basic facts on positive definite kernels and functions 
to keep the paper more self-contained. 

\subsection{Form-valued positive definite kernels} 
\mlabel{app:2}

\begin{defn} \mlabel{def:a.1} Let $X$ be a set and $V$ be a real vector space. 
We write $\Bil(V) = \Bil(V,\C)$ for the space of complex-valued 
bilinear forms on~$V$. We call a map 
$K \: X \times X \to \Bil(V)$ 
{\it a positive definite kernel} if the associated scalar-valued kernel 
\[  K^\flat \: (X \times V) \times (X \times V) \to \C, \quad 
K^\flat((x,v),(y,w)) := K(x,y)(v,w) \] 
is positive definite.\begin{footnote}
{This definition is adapted to 
our convention that scalar products are linear in the second 
argument. Accordingly, a kernel $K \: X \times X \to \Bil(V)$ is 
positive definite in the sense of Definition~\ref{def:a.1} if and only 
if the kernel $(x,y) \mapsto K(x,y)^\top$ is positive definite 
in the sense of \cite{NO15}.}
\end{footnote}

The corresponding reproducing kernel Hilbert space 
$\cH_{K^\flat} \subeq \C^{X \times V}$ is generated by the elements 
$K^\flat_{x,v}$, $x \in X, v \in V$, with the inner product 
\[   \la K^\flat_{(x,v)}, K^\flat_{(y,w)}\ra = K(x,y)(v,w) 
=: K^\flat((x,v),(y,w)) =: K^\flat_{y,w}(x,v),\] 
so that, for all $f \in \cH_{K^\flat}$, we have 
\[ f(x,v) = \la K^\flat_{x,v}, f \ra.\] 
We identify $\cH_{K^\flat}$ with a subspace of $(V^*)^X$ by identifying 
$f \in \cH_{K^\flat}$ with the function $f^* \: X \to V^*, f^*(x):= f(x,\cdot)$. 
We call 
\[ \cH_K := \{ f^* \: f \in \cH_{K^\flat} \} \subeq (V^*)^X \] 
the {\it (vector-valued) reproducing kernel space associated to $K$}. The elements 
\[ K_{x,v} := (K^\flat_{x,v})^*
\quad \mbox{ with } \quad K_{x,v}(y) = K(y,x)(\cdot, v)
\quad \mbox { for } \quad x,y \in  X, v,w \in V, \] 
then form a dense subspace of $\cH_K$ with 
\begin{equation}
  \label{eq:app}
\la K_{x,v}, K_{y,w} \ra = K(x,y)(v,w).
\end{equation}
\end{defn}

\begin{ex} If $V$ is a complex Hilbert space, $X$ is a set and 
$K \: X \times X \to B(V)$ is an operator-valued kernel, then $K$ is called 
positive definite if the corresponding kernel 
\[ \tilde K \: (X \times V) \times (X \times V) \to \C, 
\quad \tilde K((x,v),(y,w)) := \la v, K(x,y)w\ra \] 
is positive definite (\cite[Def.~I.1.1]{Ne00}), and this means that the kernel 
\[ K' \: X \times X \to \Sesq(V)\subeq \Bil(V), \quad 
K'(x,y)(v,w) := \la v, K(x,y) w \ra \] 
is positive definite. 
\end{ex}

If $X = G$ is a group and the kernel $K$ is invariant 
under right translations, then it is of the form 
$K(g,h) = \phi(gh^{-1})$ for a function 
$\phi \: G \to \Bil(V)$. 

\begin{defn} \mlabel{def:a.3} Let $G$ be a group and 
let $V$ be a real vector space.  
A function $\phi \: G \to \Bil(V)$ is said to be {\it positive definite} 
if the $\Bil(V)$-valued kernel $K(g,h) := \phi(gh^{-1})$ is positive definite. 
\end{defn}

The following proposition (\cite[Prop.~A.4]{NO15}) generalizes the GNS construction 
to form-valued positive definite functions on groups. 

\begin{prop}{\rm(GNS-construction)} \mlabel{prop:gns} Let $V$ be a real vector space.\\ 
{\rm(a)} Let $\phi \: G \to \Bil(V)$ be a positive definite function. Then 
$(U^\phi_g f)(h) := f(hg)$ defines a unitary representation of $G$ on the 
reproducing kernel Hilbert space $\cH_\phi \subeq (V^*)^G$ with 
kernel $K(g,h) = \phi(gh^{-1})$ and the range of the map 
\[ j \: V \to \cH_\phi, \quad j(v)(g)(w):=  \phi(g)(w,v), \qquad 
j(v) = K_{\1,v}^\flat, \] 
is a cyclic subspace, i.e., $U_G^\phi j(V)$ spans a dense subspace of $\cH$. 
We then have 
\begin{equation}
  \label{eq:gns-a}
\phi(g)(v,w) = \la j(v), U^\phi_g j(w) \ra 
\quad \mbox{ for } \quad g \in G, v,w, \in V.
\end{equation}

{\rm(b)} If, conversely, $(U, \cH)$ is a unitary representation of $G$ and 
$j \: V \to \cH$ a linear map whose range is cyclic, then 
\[ \phi \: G \to \Bil(V), \quad \phi(g)(v,w) := \la j(v), U_g j(w) \ra \] 
is a $\Bil(V)$-valued positive definite function and 
$(U, \cH)$ is unitarily equivalent to $(U^\phi, \cH_\phi)$. 
\end{prop}

\begin{rem} If $\phi \: G \to \Bil(V)$ is a positive definite function, 
then \eqref{eq:gns-a} shows that,if $\tilde V := \oline{j(V)}$, 
which is the real Hilbert space defined by completing $V$ with respect to the 
positive semidefinite form $\phi(\1)$, then 
\[ \tilde \phi(g)(v,w) = \la v, U_g w \ra \] 
defines a positive definite function 
\[ \tilde \phi \: G \to \Bil(\tilde V) \quad \mbox{ with } \quad 
\tilde \phi(g)(j(v),j(w)) = \phi(g)(v,w) \quad \mbox{ for } \quad v,w \in V.\] 
Therefore it often suffices to consider $\Bil(V)$-valued positive definite 
functions for real Hilbert space $V$ for which 
$\phi(\1)$ is a positive definite hermitian form on $V$ 
whose real part is the scalar product on $V$. 
In terms of \eqref{eq:gns-a}, this means that 
$j \: V \to \cH$ is an isometric embedding of the real Hilbert 
space~$V$.
\end{rem}

\subsection{Products of operator-valued kernels}

\begin{lem} \mlabel{lem:3.4} If $K_j \: X \times X \to B(V)$, $j = 1,2$, 
are two positive definite kernels with the property that 
$$ K_1(x,y) K_2(x',y') = K_2(x',y') K_1(x,y) \quad\mbox{ for } \quad 
x,x',y,y' \in X,  $$
then the product kernel $K := K_1 \cdot K_2$ is also positive definite. 
\end{lem} 

\begin{prf} Let $x_1, \ldots, x_k$. We have to show that the 
operator 
\[ C := (K_1(x_j, x_k) K_2(x_j, x_k))_{1 \leq j,k \leq n} 
\in M_n(B(V)) \cong B(V^n)\] 
is positive (cf.\ \cite[Rem.~I.1.3]{Ne00}). 

Let $\cA_j \subeq B(V)$ denote the von-Neumann algebra 
generated by the values of $K_j$. Then $\cA_1$ and $\cA_2$ commute. 
Further, the matrices 
\[ A^{(\ell)} := (K_\ell(x_j, x_k))_{1\leq j,k \leq n} \in M_n(\cA_\ell), 
\qquad \ell = 1,2, \] 
are positive, so that \cite[Lemma~4.3]{La95} implies that the matrix 
\[ D:= (K_1(x_j, x_k) \otimes K_2(x_j, x_k)) \in M_n(\cA_1 \otimes \cA_2) \] 
is positive. Since $C$ is the image of $D$ under the canonical 
representation of $M_n(\cA_1 \otimes \cA_2)$ on $V^n$, 
it follows that $C$ is positive. 
\end{prf}

\subsection{From real to complex-valued kernels} 

In this section we take a brief look at the interplay between real and complex-valued 
positive definite kernels. 
Here Corollary~\ref{cor:1.3} is of central importance because 
it shows how the positive definiteness of a complex-valued form 
$h = \gamma + i \omega$ on a real vector space $V$ leads to a 
skew-symmetric contraction on the real Hilbert space $V_\gamma$.

\begin{lem} \mlabel{lem:kernels} 
Let $K \: X \times  X\to \C$ be a positive definite kernel. 
Then the corresponding Hilbert space $\cH_K \subeq \C^X$ is invariant 
under complex conjugation such that 
$\sigma(f) := \oline f$ defines an antilinear isometry on $\cH_K$ if and only if 
$K$ is real-valued. 
\end{lem} 

\begin{prf} The invariance requirement implies the relation 
\[ \la f, K_x \ra = \oline{f(x)} = \la K_x, \sigma(f) \ra 
= \la f, \sigma(K_x) \ra \quad \mbox{ for } \quad f \in \cH_K, \] 
and therefore $\sigma(K_x) = K_x$, i.e., $K$ is real-valued. 
If, conversely, $K$ is real-valued, then 
$\cH_K = \cH_K^\R \oplus i \cH_K^\R$ is an orthogonal sum of real 
Hilbert spaces, so that complex conjugation acts on $\cH_K$ as an isometry. 
\end{prf}

\begin{prop} \mlabel{prop:1.2}  
Let $A,B \: X \times X \to \R$ be real kernels on the set $X$. 
Then the kernel 
\[ K = A + i B \: X \times X \to \C \] is positive definite 
if and only if:
\begin{itemize}
\item[\rm(a)]  $A$ is positive definite, and 
\item[\rm(b)] there exists a 
skew-symmetric contractive operator $C$ on the real reproducing kernel Hilbert space 
$\cH_A^\R \subeq \R^X$ with 
\[ B(x,y) = \la A_x, C A_y \ra= (CA_y)(x) \quad \mbox{ for } \qquad x,y \in X. \] 
\end{itemize}
\end{prop}

\begin{prf} {\bf Necessity:} If $K$ is positive definite, then so is 
$\oline K = A - iB$, and this implies that $A = \shalf(K + \oline K)$ 
is positive definite. As $A -  i B = 2 A - K$ is positive definite, 
\cite[Thm.~I.2.8]{Ne00}
\begin{footnote}{For two positive definite kernels $K$ and $Q$ on a set $X$, 
the relation $\cH_K \subeq \cH_Q$ is equivalent to 
$\lambda Q - K$ being positive definite for some $\lambda > 0$, 
and this in turn is equivalent to the existence of a bounded positive 
operator $B$ on $\cH_Q$ with $\|B\| \leq \lambda$ satisfying 
$K(x,y) = \la Q_x, B Q_y \ra = (BQ_y)(x)$ for $x,y \in X$  
(\cite[Thm.~I.2.8]{Ne00}). 
}\end{footnote}
implies the existence of a bounded operator 
$D\geq 0$ on the complex reproducing kernel Hilbert space $\cH_A \subeq \C^X$ with 
\[  K_y(x) = K(x,y) = \la A_x, D A_y \ra  = (DA_y)(x) \quad \mbox{ for } \quad 
x,y \in X.\] 
>From Lemma~\ref{lem:kernels} we know that 
$\cH_A = \cH_A^\R \oplus i \cH_A^\R$. 
>From the relation $A_y + i B_y = DA_y$ for every $y \in X$ and the fact that 
$B$ is real-valued it thus follows that 
$D = \1 + i C$ for a bounded operator $C$ on $\cH_A^\R$ satisfying 
$C A_y = B_y$ for every $y \in X$. 
Now $D = D^* \geq 0$ implies that $C = - C^\top$ is a contraction and 
\[  B(x,y) = (CA_y)(x) = \la A_x, C A_y \ra  \quad \mbox{ for } \quad 
x,y \in X.\] 

{\bf Sufficiency:} Suppose, conversely, that $A$ is positive definite and that 
$C$ is a skew-symmetric contraction on the real Hilbert space $\cH_A^\R$. 
Then the hermitian operator $\1 + i C$ on $\cH_A^\C$ is non-negative, and therefore 
its symbol 
\[ K(x,y) := \big((\1 + i C)A_y\big)(x) 
= A(x,y) + i (CA_y)(x) \] 
is a positive definite kernel on~$X$.
\end{prf}

\begin{cor} \mlabel{cor:1.3} 
Let $V$ be a real vector space, 
let $\gamma \: V \times V \to \R$ be a symmetric, 
let $\omega  \: V \times V \to \R$ be a skew-symmetric bilinear form and 
consider the corresponding hermitian form 
$h := \gamma + i \omega$. Then the following are equivalent: 
\begin{itemize}
\item[\rm(i)] $h$ is a positive definite kernel on $V$. 
\item[\rm(ii)] $\gamma$ is positive semidefinite and  
there exists a skew-symmetric bounded operator $C$ 
on the real Hilbert space $V_\gamma$ obtained by completing 
$V/\{ v \in V \: \gamma(v,v)=0\}$ such that 
$\omega(v,w) = \la  [v], C[w]\ra_{V_\gamma}$, where 
$[v]$ denotes the image of $v$ in~$V_\gamma$. 
\item[\rm(iii)] $\gamma$ is positive semidefinite and  
\begin{equation}
  \label{eq:omega-esti}
\omega(v,w)^2 \leq \gamma(v,v) \gamma(w,w) \quad \mbox{ for } \quad 
v,w \in V.
\end{equation}
\end{itemize}
\end{cor}

\begin{prf} (i) $\Leftrightarrow$ (ii): In view of Proposition~\ref{prop:1.2}, 
the kernel $h$ is positive definite if and only if the kernel 
$\gamma$ is positive definite, i.e., $\gamma$ is a positive semidefinite 
form, and the kernel $\omega$ can be written as 
\begin{equation}
  \label{eq:omega-esti2}
 \omega(v,w) = \la [v], C[w]\ra_{V_\gamma}\quad \mbox{ for } \qquad v,w \in V,
\end{equation}
where $C$ is a skew-symmetric contraction on the real Hilbert space 
$V_\gamma$. 

(ii) $\Rarrow$ (iii): \eqref{eq:omega-esti2} and $\|C\|\leq 1$ imply that 
\[ \omega(v,w)^2 \leq \|C\|^2 \|[v]\|^2 \|[w]\|^2 
= \gamma(w,w) \gamma(v,v).\] 

(iii) $\Rarrow$ (ii): 
Suppose, conversely, that $\gamma$ is positive semidefinite and that 
\eqref{eq:omega-esti} is satisfied. Then 
$\omega$ defines a continuous bilinear form on the real Hilbert space 
$V_\gamma$ with norm $\leq 1$. 
Hence there exists a skew-symmetric contraction $C \in B(V_\gamma)$ 
satisfying \eqref{eq:omega-esti2}. 
This proves the corollary.
\end{prf}

\begin{lem} \mlabel{lem:1.3} Let $h = \gamma + i \omega$ be a positive definite kernel 
as in {\rm Corollary~\ref{cor:1.3}}, let $\cH_h \subeq  \Hom(V,\C)$ 
be the corresponding reproducing kernel 
Hilbert space and let $j \: V \to \cH_h, j(v) = h(\cdot,v)$ the canonical map. 
Then the following assertions hold: 
\begin{itemize}
\item[\rm(i)] $j$ is injective if and only if $\gamma$ is positive definite, 
i.e., defines an inner product on $V$. 
\item[\rm(ii)] The complex linear 
extension $j_\C \: V_\C \to \cH_h, v + i w \mapsto j(v) + i \cdot j(w)$ 
is injective if and only if 
\[ \omega(v,w)^2 < \gamma(v,v) \gamma(w,w) \quad \mbox{ for } \quad 
0 \not=v,w \in V.\] 
\item[\rm(iii)] Suppose that 
 $\gamma$ is positive definite, that $(V,\gamma)$ is complete and that 
$\omega(v,w) = \la [v], C[w] \ra$ for an operator 
$C$ on $\cH_\gamma^\R \cong (V,\gamma)$. 
Then $j_\C$ is injective if and only if $\|Cv\| < \|v\|$ 
for every non-zero $v \in \cH_\gamma^\R$. 
\end{itemize} 
\end{lem}

\begin{prf} (i) In view of 
$\la j(v), j(w) \ra = \la h(\cdot,v), h(\cdot,w) \ra = h(v,w),$ 
we have 
$\|j(v)\|^2 = h(v,v) = \gamma(v,v),$ so 
that $j$ is injective if and only if $\gamma$ is positive definite. 

(ii) First we calculate 
\begin{align*}
\|j_\C(v+i w)\|^2 
&= \|j(v) + i \cdot j(w)\|^2 
= \gamma(v,v) + \gamma(w,w) + 2 \Re \la j(v), i \cdot j(w) \ra\\
&= \gamma(v,v) + \gamma(w,w) + 2 \Re i h(w,v) 
= \gamma(v,v) + \gamma(w,w) + 2 \omega(v,w).
\end{align*}
Writing 
$\omega(v,w) = \la \gamma_w, C\gamma_v \ra$ 
as in \eqref{eq:omega-esti2}, it follows that $j_\C(v + i w)=0$ is equivalent to 
\begin{equation}
  \label{eq:c1}
2 \la \gamma_v, C\gamma_w \ra 
= \la \gamma_v, \gamma_v \ra + \la \gamma_w, \gamma_w \ra.
\end{equation}
Next we observe that $j(v) = - i\cdot j(w)$ implies 
$\gamma(v,v) = \|j(v)\|^2 = \|j(w)\|^2 = \gamma(w,w),$ 
which leads to 
\[ \la \gamma_v, C\gamma_w \ra = \|\gamma_v\|^2 = \|\gamma_w\|^2 = \|\gamma_v\| 
\cdot \|\gamma_w\|.\] 
As $C$ is a contraction, this is equivalent to 
$C\gamma_v = \gamma_w$ by the Cauchy--Schwarz inequality.

If, conversely, there exists a non-zero $v \in V$ with 
$C\gamma_v = \gamma_w$ and $\gamma(v,v) = \gamma(w,w)$, then 
$j_\C(v+ i w) = 0$ by \eqref{eq:c1}. This proves (ii). 

(iii) If $(V,\gamma)$ is complete, $j(V) \cong (V,\gamma)$ is closed in $\cH_h$. 
Therefore $C j(V) \subeq j(V)$, and (iii) follows from the preceding discussion.
\end{prf}

\begin{rem} \mlabel{rem:1.6} If $V \subeq \cH$ is a standard real subspace 
(Definition~\ref{def:1.5}), then the kernel 
$h(v,w) := \la v,w \ra$ on $V$ has the property that the corresponding 
reproducing kernel Hilbert space is $\cH$ and the inclusion is the corresponding 
map $j \: V \to \cH$. In particular, its complex linear extension is injective. 

If, conversely, $h = \gamma + i \omega$ is a positive definite bilinear kernel 
on a real vector space $V$, then $j(V)$ is a standard real subspace of the 
corresponding complex Hilbert space $\cH_h$ if and only 
if $(V,\gamma)$ is complete (which is equivalent to the closedness of $j(V)$) and 
the complex linear extension 
$j_\C \: V_\C \to \cH_h$ is injective, which is equivalent to 
$j(V) \cap i\cdot j(V) = \{0\}$ (cf.\ Lemma~\ref{lem:1.3}(iii)). 
\end{rem}

\begin{ex} Consider the context of Proposition~\ref{prop:1.2}, where 
$K = A + i B$ is a positive definite kernel and 
$C \in B(\cH_A^\R)$ is such that $B_y = C A_y$ for $y \in X$. 
Then 
\[ V := (\1 + i C)\cH_A^\R \subeq \cH_A \] 
is a real subspace. For the isometric antilinear involution 
defined on $\cH_A$ by $\sigma(f) = \oline f$, we then have for every 
$f \in \cH_A^\R$ the relation 
\begin{align*}
 \la \sigma(\1 + i C)f, (\1 + i C)f \ra 
&= \|f\|^2 - \|Cf\|^2 \geq 0.
\end{align*}
Therefore $(\cH_A, V, \sigma)$ is a reflection positive  real Hilbert space 
(Proposition~\ref{prop:1.10}). 
\end{ex}

\subsection{Real parts of positive definite functions}

Let $\phi \: G \to \C$ be a positive definite function  
on the group $G$. Then $\oline\phi$ is also positive 
definite, so that $\Re \phi = \shalf(\phi + \oline \phi)$ 
is positive definite as well. 
>From Lemma~\ref{lem:kernels}(a) we know that a positive definite function 
$\phi$ on $G$ is real-valued if and only if the corresponding 
reproducing kernel Hilbert space $\cH_\phi$ is 
invariant under conjugation with $\|\oline f\| = \|f\|$ for 
$f \in \cH_\phi$. Based on these observations, 
one would like to understand the set of all positive definite 
functions with a given real part. A natural description of this 
set in the spirit of the present paper is provided by the following 
theorem.

\begin{thm} \mlabel{thm:1.1} {\rm(Complex extensions of real positive definite functions)} 
Let $\phi \: G \to \R$ be a positive definite function 
and let $(U^\phi, \cH_\phi^\R)$ denote the corresponding orthogonal 
representation on 
the real reproducing kernel space $\cH_\phi^\R \subeq \R^G$ 
by right translations: $(U^\phi(g)f)(h) := f(hg)$. 
Then the following assertions hold: 
\begin{itemize}
\item[\rm(a)] For each skew-symmetric contraction $C$ on $\cH_\phi$ commuting with 
$U^\phi(G)$, the 
function $\phi_C := \phi + i C \phi \in \cH_\phi \subeq \C^G$ 
is positive definite. Here we consider 
$\phi$ as an element of the real Hilbert space $\cH_\phi^\R \subeq \R^G$. 
\item[\rm(b)] Each positive definite function 
$\hat\phi$ with $\Re \hat\phi = \phi$ is of the form 
$\phi_C$ for a unique skew-symmetric contraction $C$ on $\cH_\phi$ commuting with 
$U^\phi(G)$. 
\end{itemize}
\end{thm}

\begin{prf} (a) Clearly $\cH_\phi = \cH_\phi^\R \oplus i \cH_\phi^\R$ 
is the Hilbert space complexification of $\cH_\phi^\R$ (Lemma~\ref{lem:kernels}). 
On $\cH_\phi$ the operator $B := \1 + i C$ is positive because 
it is hermitian and $\|C\| \leq 1$. Let $K(x,y) := \phi(xy^{-1})$ 
be the kernel corresponding to $\phi$ which satisfies 
$K_y = U^\phi(y)^{-1}\phi$. Then the associated kernel 
\begin{align*}
K^B(x,y) 
&:= 
\la B K_y, K_x \ra 
= \la B U^\phi(y)^{-1} \phi, U^\phi(x)^{-1} \phi \ra 
= \la U^\phi(y)^{-1} B\phi, U^\phi(x)^{-1} \phi \ra \\
&= \la U^\phi(xy^{-1}) (\1 + i C)\phi, \phi \ra 
= ((\1 + i C)\phi)(xy^{-1}) 
\end{align*}
is positive definite (cf.\ \cite[Lemma~I.2.4]{Ne00}), 
and this means that $\phi + i C \phi$ is a positive definite function. 

(b) If $\hat\phi = \phi + i \psi$ is positive definite 
with $\phi$, $\psi$ real-valued, then write 
$K = A + i B$ for the corresponding kernels: 
\[ K(x,y) = \hat\phi(xy^{-1}), \qquad 
A(x,y) = \phi(xy^{-1}) \qquad \mbox{ and } \quad 
B(x,y) =\psi(xy^{-1}).\] 
Then Proposition~\ref{prop:1.2} implies that $\phi$ is positive definite 
and that there exists a skew-symmetric contraction 
$C \in B(\cH_\phi^\R)$ with 
\[ \psi(xy^{-1}) = (CA_y)(x) = \la C U^\phi(y)^{-1} \phi, U^\phi(x)^{-1} \phi\ra.\]
Since this kernel on $G \times G$ is invariant under right translations 
and $U^\phi(G)\phi$ is total in $\cH_\phi^\R$, it follows that 
$C$ commutes with $U^\phi(G)$. This in turn leads to 
\[ \psi(xy^{-1}) 
= \la C \phi, U^\phi(yx^{-1}) \phi\ra 
= (C\phi)(xy^{-1}) \] 
and hence to $\psi = C \phi$. 
\end{prf}

\section{Standard real subspaces via contractions} 
\mlabel{sec:b}

In this section we show how standard real subspaces can be parametrized in a 
very convenient way by skew-symmetric contractions in real Hilbert spaces. 
The survey article \cite{Lo08} is an excellent source for the theory 
of standard real subspace. 

\subsection{Skew symmetric contractions} 

\begin{lem} \mlabel{lem:1.8} 
Let $C$ be a skew-symmetric contraction on the real Hilbert space 
$E$ and 
$V := (\1 + i C)E \subeq E_\C$. For $0 \not= v \in E$,  the following are equivalent: 
\begin{itemize}
\item[\rm(i)] $C^2v = - v$. 
\item[\rm(ii)] $\|Cv\| = \|v\|$. 
\item[\rm(iii)] There exists $0 \not= w \in V$ with 
$\la Cv, w \ra = \|v\| \|w\|$. 
\item[\rm(iv)] $(\1 + i C)v \in V \cap i V$. 
\end{itemize}
\end{lem}

\begin{prf} (i) $\Leftrightarrow$ (ii): First we observe that 
$\|v\|^2 - \|Cv\|^2 = \la (\1 +C^2)v,v \ra.$ 
In view of the positivity of $\1 + C^2$, the relation 
$\la (\1 + C^2)v,v \ra= 0$ is equivalent to $(\1 + C^2)v = 0$. 

(ii) $\Leftrightarrow$ (iii) follows from 
$\max \{ \la Cv, w \ra \: w \in E, \|w\| \leq 1\} = \|Cv \| \leq \|v\|.$

(iv) $\Leftrightarrow$ (i): For $w \in E$, the condition 
$(\1 + i C)v = i(\1 + i C)w$ is equivalent to 
$Cw = -v$ and $w = Cv$. Such an element $w$ exists if and only if 
$C^2v = -v$. 
\end{prf}

\begin{lem} \mlabel{lem:1.9} 
For a skew-symmetric contraction $C$ on the real Hilbert space 
$E$ and $V := (\1 + i C)E \subeq E_\C$, the following are equivalent: 
\begin{itemize}
\item[\rm(i)] $C^2 + \1$ is injective. 
\item[\rm(ii)] $\|Cv\| < \|v\|$ for every non-zero $v \in E$. 
\item[\rm(iii)] $\la Cv, w \ra < \|v\| \|w\|$ for 
non-zero elements $v,w \in E$. 
\item[\rm(iv)] $V \cap i V = \{0\}$. 
\item[\rm(v)] The operators $\1 \pm i C$ on $E_\C$ are injective. 
\item[\rm(vi)] $V + i V$ is dense in $E_\C$. 
\item[\rm(vii)] $V$ is a standard real subspace. 
\end{itemize}
\end{lem}

\begin{prf} The equivalence of (i)-(iv) follows immediately from Lemma~\ref{lem:1.8}. 

Further, (iv) can also be formulated as: $(\1 + i C)(v + i w) = 0$ for $v, w\in E$ 
implies $v + i w= 0$, which in turn means that $\1 + i C$ is injective. 
This in turn is equivalent to $\1 - i C$ being injective. Therefore (iv) is equivalent 
to (v). 

As $V + i V = (\1 + i C)E_\C = \im(\1 + i C)$, this complex subspace is dense 
if and only if the hermitian operator $\1 + i C$ has dense range, and this is equivalent to 
$\1 + i C$ being injective. Therefore (v) and (vi) are also equivalent.

Next we observe that $V$ is closed because 
\[ \|(\1 + i C)v \|^2 = \|v\|^2 + \|Cv\|^2 \geq \|v\|^2 \quad \mbox{ for } \quad v \in E \] 
shows that the range $V$ of the operator $\1 + i C \: E \to E_\C$ is closed. 
Since (iv) and (vi) are equivalent, they are therefore equivalent to $V$ being a standard 
real subspace. 
\end{prf}

\begin{prop} \mlabel{prop:1.10} Let 
$E$ be a real Hilbert space, $C$ be a skew-symmetric contraction on $E$, 
$E_\C$ be the complexification of $E$ 
and $\sigma \: E_\C \to E_\C, a + i b \mapsto a- i b$ complex conjugation on~$E_\C$. 
Then the real subspace 
\[ V := (\1 + i C)E \subeq E_\C\] 
has the following properties: 
\begin{itemize}
\item[\rm(i)] Let $E_0 = \ker(C^2 + \1)$ and $E_1 =E_0^\bot$, so that 
$E = E_0 \oplus E_1$. Then $C_0 := C\res_{E_0}$ is a complex structure 
on $E_0$ and $V_0 := (\1 + i C)E_0 \subeq E_{\C}$ is the $(-i)$-eigenspace of $C$. 
It coincides with $V \cap i V$. In particular it is a complex subspace of $E_\C$. 
The subspace $V_1 := (\1 + i C)E_1$ is a standard real subspace of $E_{1,\C}$. 
\item[\rm(ii)] If $V = V_1$, then the corresponding modular objects are given by 
$(\Delta, J) = \Big(\big(\frac{\1 - i C}{\1 + i C}\big)^2, \sigma\Big).$
\end{itemize}
\end{prop}

\begin{prf} (i) For $a, b \in E$, the relation 
$C(a + i b) = -i(a + ib)$ is equivalent to 
$C a = b$ and $Cb = -a$, i.e., to 
$a + i b \in V_0$. Therefore $V_0$ is the $(-i)$-eigenspace of $C$ in $E_\C$. 
>From Lemma~\ref{lem:1.8}(iv) we further obtain $V \cap i V = V_0$.
For $V_1 := (\1 + i C)E_1$, we thus have $V_1 \cap i V_1 = \{0\}$, 
so that Lemmas~\ref{lem:1.9}(vii) implies that 
$V_1$ is a standard real subspace of $E_{1,\C}$. 

(ii) If $V = V_1$, then 
\begin{equation}
  \label{eq:cayley}
 \Delta := \Big(\frac{\1 - i C}{\1 + i C}\Big)^2
\end{equation}
is a positive selfadjoint operator on $E_\C$ with domain 
$(\1 + i C)^2E_\C$. 
Further $\Delta^{1/2}= (\1-iC)(\1 + i C)^{-1}$ has domain 
$V_\C$

Since $\sigma\Delta \sigma = \Delta^{-1}$ by \eqref{eq:cayley}, 
$S := \sigma \Delta^{1/2}$ is an unbounded antilinear involution with 
\[  \Fix(S) = \{ \xi \in \cD(\Delta^{1/2}) = V_\C \: S \xi = \xi\}.\] 
For $\xi = (\1 + i C)v$, $v \in E_\C$, we have 
\[ S\xi = \sigma \Delta^{1/2}\xi = \sigma (\1 - i C)v = (\1 + i C) \sigma(v),\] 
so that $S\xi = \xi$ is equivalent to $v \in V$. We conclude that 
$\Fix(S) = V$. This proves (ii). 
\end{prf}

\begin{rem} Let $C$ be a skew-symmetric contraction on the real Hilbert 
space~$E$. Then the selfadjoint operator 
$C^2 + \1$ is invertible if and only if 
$-\1 \not\in \Spec(C^2)$, which is equivalent to 
$\1\not\in\Spec(iC)$, where $iC$ is considered as a selfadjoint operator 
on the complex Hilbert space~$E_\C$. 
This, in turn, is equivalent to the invertibility of $\1 + i C$ 
and hence to the boundedness of $(\1 - i C)(\1 + i C)^{-1}$. 
\end{rem}

\subsection{Real reflection positivity and standard subspaces} 

In this section we relate standard real subspaces to reflection positive 
real Hilbert spaces of the form 
$(E_\C, V, \sigma)$, where $\sigma$ is the complex conjugation on the 
complexification $E_\C$ of a real Hilbert space. This sheds an interesting
 light on the close connection between standard real subspaces 
and reflection positivity.

\begin{lem} \mlabel{lem:2.8} 
Let $E$ be a real Hilbert space and $E_\C$ be its complexification.
On $E_\C$ we consider the antilinear isometry defined by 
$\sigma(a+ i b) := a - i b$.  
A real subspace $V \subeq E_\C$ has the property that the form 
$(v,w) \mapsto \la \sigma v,w\ra$ is real-valued and positive semidefinite 
on $V$ if and only if there exists a skew-symmetric 
contraction $C \: \cD(C) \to E$ with 
$V = (\1 + i C)(\cD(C))$. 
The subspace $V$ is closed if and only if $\cD(C)$ is closed. 
\end{lem}

\begin{prf} First, let $C \: \cD(C) \to E$ be a skew-symmetric contraction 
and put $V := (\1 + i C)\cD(C)$. For $v, w \in \cD(C)$, we then have 
\begin{align*}
 \la \sigma((\1 + i C)v), (\1 + i C) w \ra 
&= \la (\1 - i C)v), (\1 + i C) w \ra 
  = \la v, w \ra + \la -iCv,w\ra + \la v, i C w \ra -  \la Cv,Cw \ra \\
&  = \la v, w \ra -  \la Cv,Cw \ra 
= \la (\1 + C^2)v,w \ra \in \R.  
\end{align*}
Moreover $\1 + C^2 \geq 0$ implies that the form is positive semidefinite. 

Conversely, let $V \subeq E_\C$ be a real subspace 
which is $\sigma$-positive in the sense that the form 
$f(v,w) := \la \sigma v,w\ra$ is real-valued and positive semidefinite. 
This assumption implies that $V \cap i E = \{0\}$. 
Hence there exists a real linear operator 
$C \: \cD(C) \to E$ for which 
$V = (\1 + i C)\cD(C)$. 
Since 
\begin{align*}
& \la \sigma(v + i C v), w + i C w \ra 
= \la v- i C v, w + i C w \ra 
= \la v, w \ra - \la Cv, C w \ra + i (\la Cv, w \ra + \la Cw, v \ra)  
\end{align*}
is supposed to be real-valued, 
\[ \la Cv, w \ra + \la v, C w\ra=0 \quad \mbox{ for }  \quad 
v,w \in E.\] 
This means that $C$ is skew-symmetric on $\cD(C)$. 
Further, the positivity assumption implies that 
$\|Cv\| \leq \|v\|$ for $v \in E$. 

The subspace $V$ is closed if and only if the graph of $C$ is closed, 
which is equivalent to the closedness of $\cD(C)$ because 
$C$ is a contraction. 
\end{prf}

\begin{prop} \mlabel{prop:1.10b} Let 
$E$ be a real Hilbert space, $C$ be a skew-symmetric contraction on $E$, 
$E_\C$ be the complexification of $E$ 
and $\sigma \: E_\C \to E_\C, a + i b \mapsto a- i b$ complex conjugation on~$E_\C$. 
Then the real subspace 
\[ V := (\1 + i C)E \subeq E_\C\] 
has the following properties: 
\begin{itemize}
\item[\rm(i)] $V$ is closed and $\sigma$-positive, so that 
$(E_\C, V, \sigma)$ is a reflection positive real Hilbert space. 
\item[\rm(ii)] $V^\bot = i \sigma(V)$, i.e., the bilinear 
form $\gamma_\sigma(\xi,\eta) := \la \sigma \xi,\eta\ra$ on $V$ is real-valued. 
\item[\rm(iii)] The null space of the positive semidefinite 
form $\gamma_\sigma$ on $V$ 
coincides with the $(-i)$-eigenspace $V_0$ of $C$ on $E_\C$. 
If $V_0 = \{0\}$, then the unbounded positive operator 
\[ F := \sqrt{\frac{\1 - i C}{\1 + i C}} \: V \to E_\C \] 
satisfies  $\|F\xi\|^2 = \la \sigma\xi,\xi\ra$ for $\xi \in V$, 
so that we can identify the real Hilbert space completion 
$\hat V$ of $V$ with respect to $\gamma_\sigma$ with 
$\oline{F(V)}$. We further have 
$\sigma F \sigma = F^{-1}$.
\end{itemize}
\end{prop}

\begin{prf} (i) The subspace $V$ is closed because 
\[ \|(\1 + i C)v \|^2 = \|v\|^2 + \|Cv\|^2 \geq \|v\|^2 \quad \mbox{ for } \quad v \in E \] 
shows that the range of the operator $\1 + i C \: E \to V$ is closed. 

For the complex conjugation $\sigma$ on $E_\C$,  we have for $v,w \in E$ the relation 
\begin{align*}
\gamma_\sigma((\1 + i C)v, (\1 + i C)w)
&= \la \sigma(\1 + i C)v, (\1 + i C)w \ra 
=  \la (\1 - i C)v, (\1 + i C)w \ra \\
&=  \la (\1 + i C)(\1 - i C)v, w \ra=  \la (\1 + C^2)v, w \ra  \in \R
\end{align*}
and thus 
\[ \gamma_\sigma((\1 + i C)v, (\1 + i C)v) = \|v\|^2 - \|Cv\|^2 \geq 0.\] 

(iii) An element $a + i b \in E_\C$ ($a,b \in E$) 
is orthogonal to $V$ with respect to the real scalar product if and only if 
\[ 0 = \Re \la a + i b, v + i C v \ra = \la a,v \ra + \la b, C v \ra 
= \la a - Cb,v \ra \] 
for every $v \in E$, and this is equivalent to $Cb = a$, i.e., to 
$a + i b = i(b - i Cb) \in i \sigma(V)$. 


(iv) An element $\xi := (\1 + i C)v \in V$ satisfies 
$\la \sigma \xi ,\xi \ra = 0$ if and only if $C^2 v = - v$, 
which is equivalent to 
\[ (\1 - i C)\xi = (\1 - i C)(\1 + i C)v = (\1 + C^2)v = 0,\] 
i.e., to $C\xi = - i\xi$. This implies that $V_0 \subeq V$ is the nullspace 
of $\gamma_\sigma$. 

Now we assume that $V_0 = \{0\}$ and $V = V_1$. 
As $\1 \pm i C$ are non-negative hermitian operators on $E_\C$, they have 
a non-negative square root and $(\1 + i C)^{-1/2}$ is an unbounded 
operator whose domain is 
\[ \sqrt{\1 + i C}E_\C \supeq 
\sqrt{\1 + i C}\sqrt{\1 + i C} E_\C = (\1 + i C)E_\C.\]
This leads to an unbounded symmetric operator 
\[ F := \sqrt{\frac{\1 - i C}{\1 + i C}} \: V \to E_\C. \] 
For $\xi = (\1 + i C)v$, $v \in E$, we have 
\[ F\xi = \sqrt{(\1 - i C)(\1 + i C)} v = \sqrt{\1 + C^2} v,\] 
so that 
$\|F\xi\|^2 = \la (\1 + C^2)v, v \ra = \la \sigma\xi,\xi \ra.$ 
Therefore $F \: V \to \hat V := \oline{F(V)} \subeq E_\C$ is the canonical map 
of the reflection positive real Hilbert space 
$(E_\C, V, \sigma)$. It satisfies 
\[ \sigma F \sigma = \sqrt{\frac{\1 + i C}{\1 - i C}} = F^{-1}.\qedhere\] 
\end{prf}

\begin{rem} Since $U_t = \Delta^{-it}$ acts on 
the reflection positive Hilbert space $(E_\C, V, \sigma)$ 
by automorphisms, it induces on the corresponding real Hilbert space $\hat V$ 
an orthogonal representation. The natural map 
$\sqrt{\1 + C^2} \: E \to \hat V$ 
in Proposition~\ref{prop:1.10b} intertwines the orthogonal representations 
$U_t\res_E$ and $U_t\res_{\hat V}$. 
\end{rem}

The following proposition asserts that all standard real subspace are of the 
form described in Proposition~\ref{prop:1.10}.

\begin{prop} \mlabel{prop:2.9} 
Let $V \subeq \cH$ be a standard real subspace with 
modular objects $(\Delta, J)$. Then 
$E := \Fix(J)$ is a real Hilbert space with $\cH \cong E_\C$ and there 
exists a skew-symmetric strict contraction $C \: E \to E$ with 
$V = (\1 + i C)E$. Then $\cD(\Delta) \cap V$ is dense in~$V$.
\end{prop}

\begin{prf} First we observe that $V$ is $J$-positive: 
\[ \la J\xi, \xi \ra = \la J S\xi, \xi \ra = \la \Delta^{1/2}\xi,\xi \ra \geq 0.\] 
This implies the existence of a contraction 
$C \: \cD(C) \to E$ with 
\[ V = \Gamma(C) := (\1 + i C)\cD(C)\]  
(Subsection~\ref{subsec:3.2}). 
That $C$ is strict follows from Lemma~\ref{lem:1.9}. 
>From the real orthogonal decomposition 
$\cH = V \oplus i J(V)$  (\cite[Lemma~4.2(iv)]{NO15}) we now obtain 
\[ V^\bot = i J(V) = i (\1 - i C)\cD(C) = i \Gamma(-C) = (C + i \1)\cD(C), \] 
where $\bot$ refers to the real-valued scalar product $\Re \la \cdot,\cdot \ra$ 
on $\cH \cong E \oplus i E$. 

If $a \in E \cap \cD(C)^\bot$, then $a \in V^\bot = i J(V) = i \Gamma(-C)$ leads to $a = C0 = 0$. 
Therefore $\cD(C)$ is dense in~$E$. As $V$ is closed and 
$\1 + i C \: \cD(C) \to V$ is a topological isomorphism, it follows that 
$\cD(C)$ is closed, and thus $\cD(C) = E$. 

As $\gamma_J(\xi,\eta) := \la J\xi,\eta\ra$ is real-valued on $V$ 
(recall $JV = (iV)^\bot$), we obtain 
for $v,w \in V$ the relation 
\begin{align*}
0 &= \Im \la J(\1 + i C)v, (\1 + i C)w \ra 
= \Im \la (\1 - i C)v, (\1 + i C)w \ra \\
&= \Im \la (\1 - i C^\top)(\1 - i C)v, w \ra 
= - \la (C^\top + C)v,w \ra,
\end{align*}
so that $C^\top = - C$ (Lemma~\ref{lem:2.8}). 

It remains to show that 
$\cD(\Delta) \cap V$ is dense in $V$. 
Since $C$ is a strict contraction, the kernel of $\1 + C^2$ is trivial, 
resp., $-1$ is not an eigenvalue of $C^2$. Let 
$E_n \subeq E$ be the spectral subspace of $C^2$ for the subset 
$[-1+1/n,1]$. This subspace is $C$-invariant and the union of these 
subspace is dense in $E$ because $-1$ is not an eigenvalue. 
As $(\1 + i C)E_n \subeq \cD(\Delta)$, it follows that 
$\cD(\Delta) \cap V$ is dense in $V$. 
\end{prf}

\subsection{Contractions and modular objects} 

The following lemma describes the complex-valued scalar product on a standard 
real subspace in terms of the corresponding modular objects $(\Delta, J)$. 

\begin{lem} \mlabel{lem:6.10}
Let $V \subeq \cH$ be a standard real subspace, $(\Delta, J)$ be the 
corresponding modular objects and 
\[ \la v,w\ra_\cH = \gamma(v,w) + i \omega(v,w) \] 
be the corresponding hermitian positive definite form on $V$;  in particular 
$\la v,w \ra_V = \gamma(v,w)$. Then 
\begin{equation}
  \label{eq:realpart}
\gamma(v,w) = \frac{1}{2}\Big(\la v, w \ra + \la \Delta^{1/2}v, \Delta^{1/2}w \ra\Big)
\quad \mbox{ and }\quad 
\omega(v,w) 
= \frac{1}{2i}\Big(\la v, w \ra - \la \Delta^{1/2}v, \Delta^{1/2}w \ra\Big). 
\end{equation}
In particular, we have a strict contraction $C$ on $V$ satisfying: 
\begin{equation}
  \label{eq:C-rel}
 \omega(v,w) = \gamma(v,Cw) \quad \mbox{ and } \quad 
C = \hat C\res_V, \quad \hat C = i \frac{\Delta - \1}{\Delta + \1} = i \frac{\Delta^{1/2} - \Delta^{-1/2}}
{\Delta^{1/2} + \Delta^{-1/2}} = i \tanh\Big(\frac{\log \Delta}{2}\Big).
\end{equation}
It satisfies 
\begin{equation}
  \label{eq:h-rel}
\la v,w \ra_\cH = \la v,  (\1 + i C)w \ra_{V_\C} \quad \mbox{ for } \quad 
v,w \in V_\C, 
\end{equation}
so that the map 
$\Phi := \sqrt{\1 + i C} \: V_\C \to V_\C$ 
extends to an isometric inclusion $\cH \into V_\C$. 
\end{lem}

\begin{prf} As $V \subeq \cD(\Delta^{1/2})$ and 
$v = S v = J \Delta^{1/2}v$ or $v \in V$ (Remark~\ref{rem:2.2b}), we obtain 
\[ \la \Delta^{1/2}v, \Delta^{1/2}w \ra = \la Jv, Jw \ra = \la w,v \ra = \oline{\la v, w \ra} 
\quad \mbox{ for } \quad v,w \in V.\] 
This implies \eqref{eq:realpart}. 
Next we note that $B := \frac{\Delta - \1}{\Delta + \1}$ is a bounded 
operator on $\cH$ which can also be written as 
\[ B = \frac{\Delta^{1/2} - \Delta^{-1/2}}{\Delta^{1/2} + \Delta^{-1/2}}.\] 
In this form we see that $JBJ = - B$. We also note that $B$ commutes 
with $\Delta$, hence preserves $\cD(\Delta^{1/2})$. This leads to 
\[ S B = J \Delta^{1/2} B = - B S,\] 
and therefore to $BV = B \Fix(S) \subeq  i \Fix(S) = i V$. In particular, 
$\hat C := i B$ restricts to a bounded skew-symmetric operator $C \: V \to V$. 
If $v,w$ are contained in the dense subspace $V \cap \cD(\Delta)$ of $V$ 
(Proposition~\ref{prop:2.9}), 
we obtain
\begin{align*}
 \gamma(v,Cw) 
&= \frac{1}{2}\Big(\la v, Cw \ra + \la \Delta^{1/2}v, \Delta^{1/2}Cw \ra\Big) 
= \frac{1}{2}\la (\1 + \Delta)v, Cw \ra \\ 
&= \frac{1}{2}\la v, (\1 + \Delta)\hat C w \ra 
= \frac{1}{2i}\la v, (\1-\Delta)w \ra = \omega(v,w).
\end{align*}
Since $\omega$ and $\gamma(\cdot,C\cdot)$ are continuous on $V$, they 
coincide on all of $V$. By Lemma~\ref{lem:1.9}, the operator $C$ is a strict contraction. 
By \eqref{eq:C-rel}, we have for $v,w \in V$ the relation 
\eqref{eq:h-rel}, and since both sides are sesquilinear, it also holds 
for $v,w \in V_\C$. 
\end{prf}

\end{document}